\documentclass[aps,prd,reprint,floatfix,superscriptaddress,lengthcheck,sort&compress,letterpaper,nofootinbib]{revtex4-1}

\usepackage{graphicx}% Include figure files
\usepackage{dcolumn}% Align table columns on decimal point
\usepackage{bm}% bold math
\usepackage{dsfont}
\usepackage{amsmath}
\usepackage[usenames]{color}
\usepackage[colorlinks=true,linkcolor=blue,citecolor=blue,urlcolor=blue]{hyperref}

\newcommand{\gS}[1]{#1\!\!\!\!\!\not~}

\newcommand{\pslash}{\gS{p}}

\begin{document}

\title{The $\sigma$-meson: four-quark vs. two-quark components and decay width in a Bethe-Salpeter approach}

\author{Nico Santowsky}
 \email[e-mail: ]{nico.santowsky@theo.physik.uni-giessen.de}
\affiliation{Institut f\"ur Theoretische Physik, Justus-Liebig Universit\"at Gie{\ss}en, 35392 Gie{\ss}en, Germany}
\author{Gernot Eichmann}
\email[e-mail: ]{gernot.eichmann@tecnico.ulisboa.pt}
\affiliation{LIP Lisboa, Av.~Prof.~Gama~Pinto 2, 1649-003 Lisboa, Portugal}
\affiliation{Departamento de F\'isica, Instituto Superior T\'ecnico, 1049-001 Lisboa, Portugal}
\author{Christian S. Fischer}
 \email[e-mail: ]{christian.fischer@theo.physik.uni-giessen.de}
\affiliation{Institut f\"ur Theoretische Physik, Justus-Liebig Universit\"at Gie{\ss}en, 35392 Gie{\ss}en, Germany}
\affiliation{Helmholtz Forschungsakademie Hessen f\"ur FAIR (HFHF),
	GSI Helmholtzzentrum f\"ur Schwerionenforschung, Campus Gie{\ss}en, 35392 Gie{\ss}en, Germany}
\author{Paul C. Wallbott}
\affiliation{Institut f\"ur Theoretische Physik, Justus-Liebig Universit\"at Gie{\ss}en, 35392 Gie{\ss}en, Germany}
\author{Richard Williams}
\affiliation{Institut f\"ur Theoretische Physik, Justus-Liebig Universit\"at Gie{\ss}en, 35392 Gie{\ss}en, Germany}

\date{\today}

\begin{abstract}
We study the dynamical generation of resonances in isospin singlet channels with mixing between two- and four-quark states. To this end we
generalise a Bethe-Salpeter approach to four-quark states employed previously to accommodate for mixing diagrams.
The $q\bar{q}q\bar{q}$ and $q\bar{q}$ components of the Bethe-Salpeter wave function (with light quarks $q\in\{u,d\}$) are determined
consistently in a symmetry-preserving truncation of the underlying Dyson-Schwinger equations.
As a prominent example we deal with the isospin-singlet $0^{++}$ meson with light quark content.
We find that the $\pi\pi$ contribution of the four-quark component is mainly responsible for the low (real part of the)
mass of the resulting state. We also study the analytic structure in the complex momentum plane and find a branch cut at the two-pion threshold
and a singularity in the second Riemann sheet indicating a considerable decay width. Our findings are in excellent qualitative agreement with the
general picture for the $\sigma/f_0(500)$ that emerged in the past two decades from dispersive approaches.
\end{abstract}

\maketitle

%\tableofcontents

\section{\label{sec:1}Introduction}

There is perhaps no other state in the low-energy spectrum of QCD that has been puzzled over so intensely over the past decades than the isoscalar-scalar
meson, the $f_0(500)$ or $\sigma$ meson. It plays a very important role in our understanding of chiral symmetry breaking, it is one of the defining
quantities in model building and, together with its flavor partners in the light scalar nonet, it has properties that do not go easily together
with a conventional mesonic $q\bar{q}$ picture of its internal structure. Once discarded from the particle data book, the light scalar meson nonet
has been reintroduced again at the end of the past century. Only in the 2012 edition the $\sigma$ meson has been renamed $f_0(500)$ and accepted with
much smaller uncertainties in its mass and width as ever before, mainly due to progress in the dispersive analysis of experimental data, see
\cite{Caprini:2005zr,Yndurain:2007qm,GarciaMartin:2011jx,Moussallam:2011zg} and Refs. therein.
Many aspects of this fascinating story together with a detailed
discussion on the theoretical background and the intricacies of data analysis can be found in the review article Ref.~\cite{Pelaez:2015qba}.

The notion that the multiplet of light scalar mesons is incompatible with a conventional $q\bar{q}$ picture goes back some way \cite{Jaffe:1976ig}:
by assuming a dominant four-quark structure, interesting properties like inverted mass hierarchies and decay patterns are naturally explained.
This picture is supported by effective theory studies and large-$N_c$ arguments (see e.g.
\cite{Achasov:1987ts,Black:1998wt,Maiani:2004uc,Giacosa:2006rg,Klempt:2007cp,Ebert:2008id,Pelaez:2015qba} and references therein) as well as
lattice calculations~\cite{Alford:2003xw,Mathur:2006bs,Prelovsek:2010ty,Prelovsek:2010kg} and functional methods \cite{Heupel:2012ua,Eichmann:2015cra}.

In principle, an experimentally observed state may be a mixture of several states with different internal structure but the same quantum numbers.
Thus, it is by no means clear whether the $f_0(500)$ is indeed a pure four-quark state or has some overlap with the corresponding
conventional $q\bar{q}$ state, thus generating subdominant $q\bar{q}$-contributions to its composition. This idea has been elaborated upon
in a number of works including studies within linear sigma models \cite{Black:1998wt,Close:2002zu,Giacosa:2006tf}, instanton induced mixings in 
tetraquark models \cite{Hooft:2008we}, unitarized quark model calculations \cite{Londergan:2013dza} and unitarized chiral perturbation theory 
and leading $1/N_c$-considerations \cite{Pelaez:2006nj,Nieves:2011gb,RuizdeElvira:2010cs}. Evidence from these studies clearly points towards 
a small but significant admixture of $q\bar{q}$-components to the four-quark state.
Note that a similar situation also appears in the heavy-quark sector, where several XYZ states are discussed to be dominated
by four-quark components with a heavy and a light quark-antiquark pair. However, at least for the isospin-singlet cases like the $\chi_{c1}(3872)$,
substantial admixtures of $\bar{c}c$ components seem possible \cite{Prelovsek:2013cra}.

In this work we address this question on the example of the isospin-singlet $0^{++}$ meson with light quark content using the functional
approach via Dyson-Schwinger (DSE) and Bethe-Salpeter equations (BSE). In previous work within this approach, the state in question has been treated
either as a conventional $q\bar{q}$ state using a range of truncations for the quark-gluon interaction, or as a pure four-body $q\bar{q}q\bar{q}$ state.
Using the rainbow-ladder approach in the $q\bar{q}$ BSE, this state is found at a (real) mass of the order of 650 MeV \cite{Cotanch:2002vj}. In
various beyond rainbow-ladder calculations with sophisticated interactions its (real) mass is shifted to (much) larger values beyond 1 GeV
\cite{Chang:2009zb,Fischer:2009jm,Williams:2015cvx}. This does not support a large quark-antiquark contribution to the $f_0(500)$.
On the other hand, the treatment of the lightest scalar meson as a four-quark state, once in the four-body approach using a Faddeev-Yakubovsky
equation \cite{Eichmann:2015cra} and once in a reduced two-body approach \cite{Heupel:2012ua} using internal meson and diquark degrees of freedom,
led to substantially smaller masses around 400 MeV. Moreover, it turned out that the latter two approaches both favoured a leading internal $\pi\pi$
component in the state's wave function, thus suggesting the identification with the $f_0(500)$.

While these studies are already suggestive, they can be improved in two respects. First, it needs to be studied whether the picture of $\pi\pi$
dominance still persists when the coupled system of Bethe-Salpeter equations for the four-quark $q\bar{q}q\bar{q}$ and conventional $q\bar{q}$
states is considered. Second, previous extractions of the $\sigma$ mass in the four-quark formalism focused on the real part only and were not able
to determine the width of the state in question. We will address both problems in this work.

The paper is organised as follows: We first detail our derivation of the coupled system of BSEs for a state with four-quark
$q\bar{q}q\bar{q}$ and two-quark $q\bar{q}$ components in section \ref{sec:2}. To keep matters as simple as possible, we use the two-body
approximation of the four-body equation introduced in \cite{Heupel:2012ua}. In section \ref{sec:3.1} we then detail our model for the underlying
quark-gluon interaction and discuss briefly the corresponding DSE for the quark propagator. Together, the resulting quark
and the interaction provide a self-consistent input into the various BSEs studied in this work. In section \ref{sec:3.2} we summarise
an approach to deal with the analytic structure generated by the two-pion cut, developed in Ref.~\cite{Williams:2018adr} and adapted here for the
case of the $\sigma$ meson. Section \ref{sec:4} is devoted to a discussion of our results. In \ref{sec:4.1} we discuss our results for the
coupled system of four- and two-quark BSEs and study the impact of mixing effects. In section \ref{sec:4.2} we then determine the decay
width of the pure two-quark component. We summarise and conclude in section \ref{sec:5}.

\section{\label{sec:2}The functional approach\label{thedsebseapproach}}

\subsection{\label{sec:2.1}The four-body Bethe-Salpeter equation \label{dsesandbses}}

The equations of motion for $n$-quark bound states, the Bethe-Salpeter equations (BSEs) are derived from the $(2n)$-quark scattering matrix $T^{(n)}$
and the corresponding scattering kernel $K^{(n)}$:
\begin{equation}
	T^{(n)}=K^{(n)}+K^{(n)}G_0^{(n)}T^{(n)}\,.\label{eq-dysonsum}
\end{equation}
Here $G_0^{(n)}$ denotes the product of $n$ dressed quark propagators. By defining a bound-state amplitude (BSA) $\Gamma^{(n)}$, which carries
the Dirac, color and flavor structure of the bound state in question, the $T$ matrix at the physical pole $P^2 \rightarrow -M^2$
of the bound state propagator $D$ is given by
\begin{equation}
	T^{(n)} \overset{P^2 \rightarrow -M^2}{\approx}  \frac{\Gamma\bar\Gamma}{P^2+M^2}\,.\label{eq-bsaansatz}
\end{equation}
Inserting this expression into Eq.~(\ref{eq-dysonsum}) leads to the homogeneous on-shell $n$-quark BSE:
\begin{equation}
	\Gamma^{(n)}=K^{(n)}G_0^{(n)}\Gamma^{(n)} \,.\label{eq-bsegeneral}
\end{equation}
In this paper we focus on 2-quark and 4-quark bound states and we further denote $\Gamma^{(2)} \equiv \Gamma$ as a two-quark BSA
and $\Gamma^{(4)} \equiv \Psi$ as a four-quark BSA.

\begin{figure}
	\includegraphics[width=8.8cm]{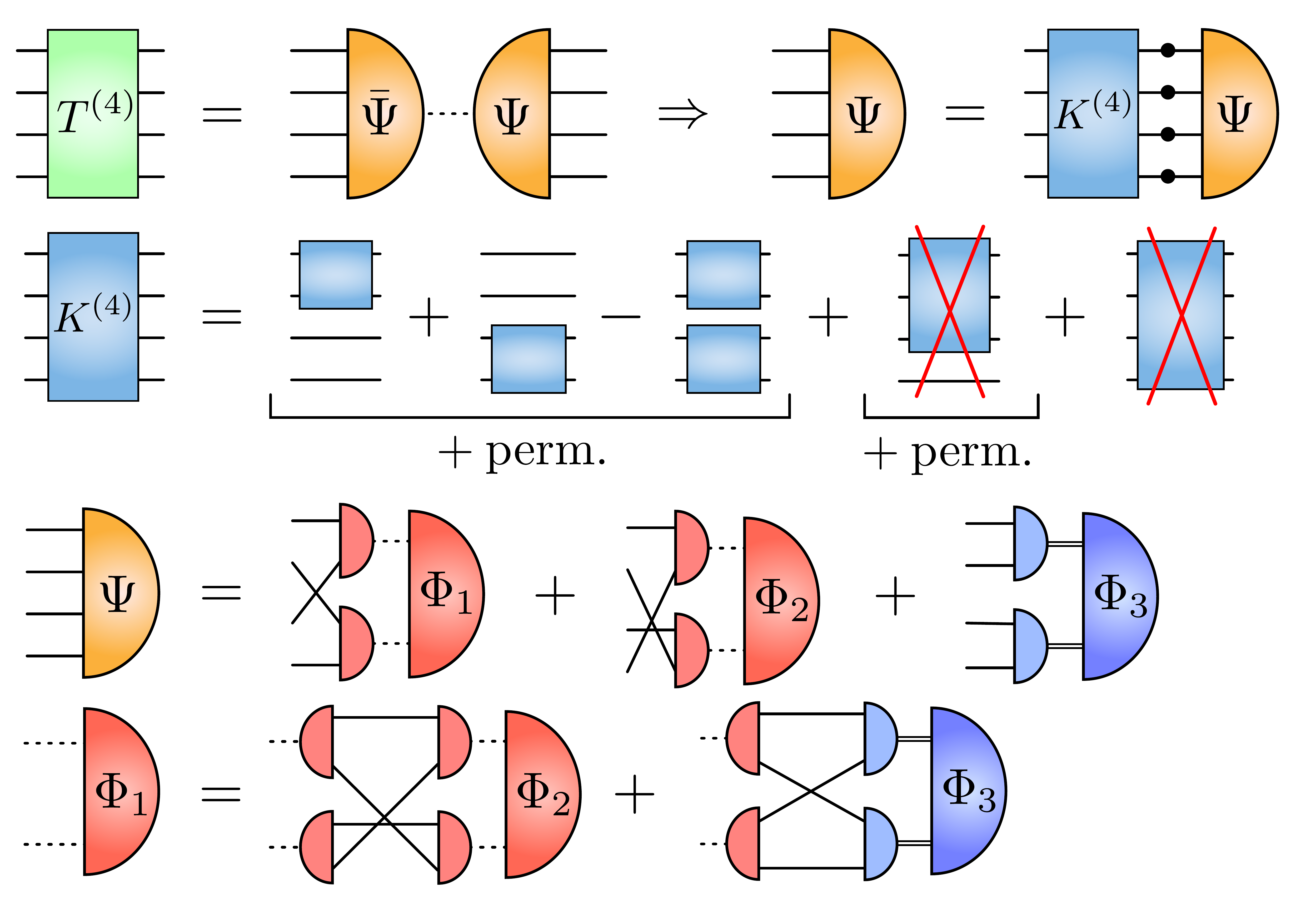}
	\caption{Diagrammatic representation of the basic quantities used in deriving the pure two-body/four-quark BSE.
		In the first line we display the representation of the bound state (\ref{eq-bsaansatz}) together with its corresponding BSE (\ref{eq-bsegeneral}).
		In the second line we give the explicit decomposition of the interaction kernel $K^{(4)}$ in terms of irreducible two-, three- and
		four-body interactions. The red crosses indicate truncations, justified and explained in the main text.
		The third line displays the reduction of the four-body amplitude into a sum
		of two-body amplitudes featuring internal mesons (dashed lines) and \mbox{(anti-)}diquarks (double lines). One of the resulting effective two-body
		equations is given in the lowest line. The other two equations are obtained under permutations in the index set $\{1,2,3\}$, thus spanning
		the whole coupled system of two-body BSEs (for equal quark masses $\Phi_1$ and $\Phi_2$ are identical).\label{fig-diagram1}}
\end{figure}
The complete scattering kernel $K^{(4)}$ occurring in the four-body BSE,
\begin{align}
	\Psi    &   = K^{(4)}\:G_0^{(4)}\:\Psi,\label{BSA}\\
	K^{(4)}	&	=\tilde{K}^{(2)}+\tilde{K}^{(3)}+\tilde{K}^{(4)}
\end{align}
can be decomposed in three contributions $\tilde{K}^{(2,3,4)}$ containing irreducible two-, three- and four-quark interactions \cite{Khvedelidze:1991qb}.
In the following we adopt the strategy of Ref.~\cite{Heupel:2012ua,Eichmann:2015cra} and neglect the latter two, i.e. we set
$\tilde{K}^{(3)}=\tilde{K}^{(4)}=0$. The basic idea behind this approximation is the notion that the two-body interactions
may well dominate the state in question. This is motivated by two considerations: first, the broad decay width of the experimental $f_0(500)$ into two pions
suggests a dominance of internal two-body correlations. Conversely, if strong
quark-antiquark correlations inside the four-quark state form a two-pion internal structure, its singularity structure naturally
has a strong effect on the BSE that almost inevitably will dominate the equation and dwarf contributions from $\tilde{K}^{(3)}$ and
$\tilde{K}^{(4)}$. Second, a similar approximation has been applied with great success in the baryon sector. There, strong two-body
correlations naturally lead to a diquark-quark picture, which in turn leads to a spectrum in one-to-one agreement with
experiment, see e.g. \cite{Eichmann:2016hgl,Eichmann:2016yit} and references therein. Moreover, for baryons it can be shown explicitly 
that the leading part of the irreducible three-body interaction (in terms of a skeleton expansion) is small \cite{Sanchis-Alepuz:2017jjd}.
While all these arguments are not strictly rigorous, they provide plausible justification for neglecting the irreducible many-body
interactions $\tilde{K}^{(3,4)}$ also in the four-quark case. Practicability arguments complement these considerations.

The contribution $\tilde{K}^{(2)}$ containing all irreducible two-body interactions inside the four-quark state contains various incarnations
of the two-body scattering kernel ${K}^{(2)}_{ij}$ between two quarks $i$ and $j$:
\begin{align}
	\tilde{K}^{(2)}	&	=\underbrace{{K}^{(2)}_{12}S^{-1}_3S^{-1}_4+{K}^{(2)}_{34}S^{-1}_1S^{-1}_2
		                                   -{K}^{(2)}_{12}{K}^{(2)}_{34}}_{=:\tilde{K}^{(2)}_{(12)(34)}}
	                    + \text{perm.}\nonumber\\
	                &   = \sum_a \tilde{K}^{(2)}_a \label{eq-2bodykernels}
\end{align}
Explicit indices $1,2,3,4$ denote the four (anti-)quarks as ingredients of the four-quark bound state and the summation over $a$ picks up the
three possible combinations $(12)(34), (13)(24), (14)(23)$ of two-body interactions.

\subsection{\label{sec:2.2}The two-body Bethe-Salpeter equation}

In order to be able to extract a two-body Bethe-Salpeter equation for the four-quark $q\bar{q}q\bar{q}$ state, we slightly reformulate the problem
 \cite{Heupel:2012ua}. First, we define a four-body T-matrix $T_a$ that is generated by $\tilde{K}^{(2)}_a$:
\begin{equation}
T_a = \tilde{K}^{(2)}_a + \tilde{K}^{(2)}_a G_0^{(4)} T_a = \tilde{K}^{(2)}_a +  T_a G_0^{(4)} \tilde{K}^{(2)}_a\,. \label{Ta}
\end{equation}
Furthermore, we note that the BSA, Eq.~(\ref{BSA}), can be split into three separate parts by inserting Eq.~(\ref{eq-2bodykernels})
\begin{align}
\Psi = \sum_a \tilde{K}^{(2)}_a G_0^{(4)}\:\Psi := \sum_a \Psi_a\,.
\end{align}
Acting with $T_a G_0^{(4)}$ onto $\Psi$ and using (\ref{Ta}) one then obtains
\begin{equation}
\Psi_a=T_a \,G_0^{(4)}\,(\Psi-\Psi_a)=\sum_{b\neq a}\,T_a\,G_0^{(4)}\,\Psi_b\,, \label{eq-masterbse}
\end{equation}
which is still an exact four-body equation apart from neglecting the kernels $\tilde{K}^{(3)}$ and $\tilde{K}^{(4)}$.

Since the T-matrices $T_a$ contain effects from two-body interactions in the same combination of (anti-)quark legs only, they are prone to
develop singularities in the respective channels, with the quantum numbers of mesons and (anti-)diquarks. The two-body approximation of the
four-body equation then amounts to replacing $T_a$ with a pole ansatz analogously to Eq.~(\ref{eq-bsaansatz}). Assuming that the spin-momentum
structure of the Bethe-Salpeter amplitudes factorizes, the full amplitude $\Psi$ can then be decomposed into meson-meson and diquark-antidiquark
substructures $\Phi_{a}$. We thus obtain
\begin{equation}
	\Psi_a = \left(\Gamma_{12}\otimes\Gamma_{34}\right)\:G_0^{(2,2)}\:\Phi_{a} \label{eq-4bsastructure}
\end{equation}
for $a=(12)(34)$ and similar expressions for the other combinations. Here $G_0^{(2,2)}$ is a combination of two meson propagators or a diquark and an
antidiquark propagator, respectively, and $\Gamma_{ij}$ are the corresponding two-body Bethe-Salpeter amplitudes. The representation
Eq.~(\ref{eq-4bsastructure}) is in some sense a `physical basis' in that it builds a representation of $\Psi_a$ in terms of reduced internal Dirac,
flavor and color structure from a physical picture. The algebraic structure
of the tetraquark-meson and tetraquark-diquark vertices $\Phi_a$ depend on the respective quantum numbers of the investigated four-quark state.
For scalar four-quark states and (pseudo)scalar ingredients, e.g., those amplitudes are flavor and color singlets and Lorentz scalars.

With Eq.~(\ref{eq-4bsastructure}), we effectively solve for the vertices $\Phi_a$ while making use of solutions of the two-quark
BSE for the amplitudes $\Gamma_{ij}$. The interaction kernel elements for the internal vertices $\Phi_a$ are quark exchange diagrams
as visualized in the last line of Fig. \ref{fig-diagram1}.

\subsection{\label{sec:2.3}Inclusion of a two-quark component \label{ssec-inclusiontwoquark}}

We now extend the truncation for the four-body equation (\ref{eq-masterbse}) by adding a phenomenologically motivated two-quark component
into the Bethe-Salpeter amplitude. For $a=(12)(34)$ in Eq.~(\ref{eq-4bsastructure}) this amounts to
\begin{equation}
\Psi_a = \left(\Gamma_{12}\otimes\Gamma_{34}\right)\:G_0^{(2,2)}\:\Phi_{a} +
     K^{(2)}_{13} K^{(2)}_{24} S_{34} G_0^{(2)}\Gamma^*_{12}  +\text{perm.}
     \label{eq-4bsastructure_v2}
\end{equation}
where $\Gamma^*_{12}$ is the Bethe-Salpeter amplitude of a quarkonium state with the same quantum numbers as the four-body state connected
to the propagator lines $1$ and $2$, and the quark propagator $S_{34}$ connects the lines $3$ and $4$.
This is equivalent to extending the physical basis discussed above with another possible basis element. Note that these extensions only
appear in those $\Psi_a$ with two-body interactions between quark-antiquark pairs, i.e. meson-meson contributions. They do not appear in
the $\Psi_a$ with two-body interactions between quark-quark pairs, i.e. in diquark/antidiquark contributions.

\begin{figure}
	\includegraphics[width=8.7cm]{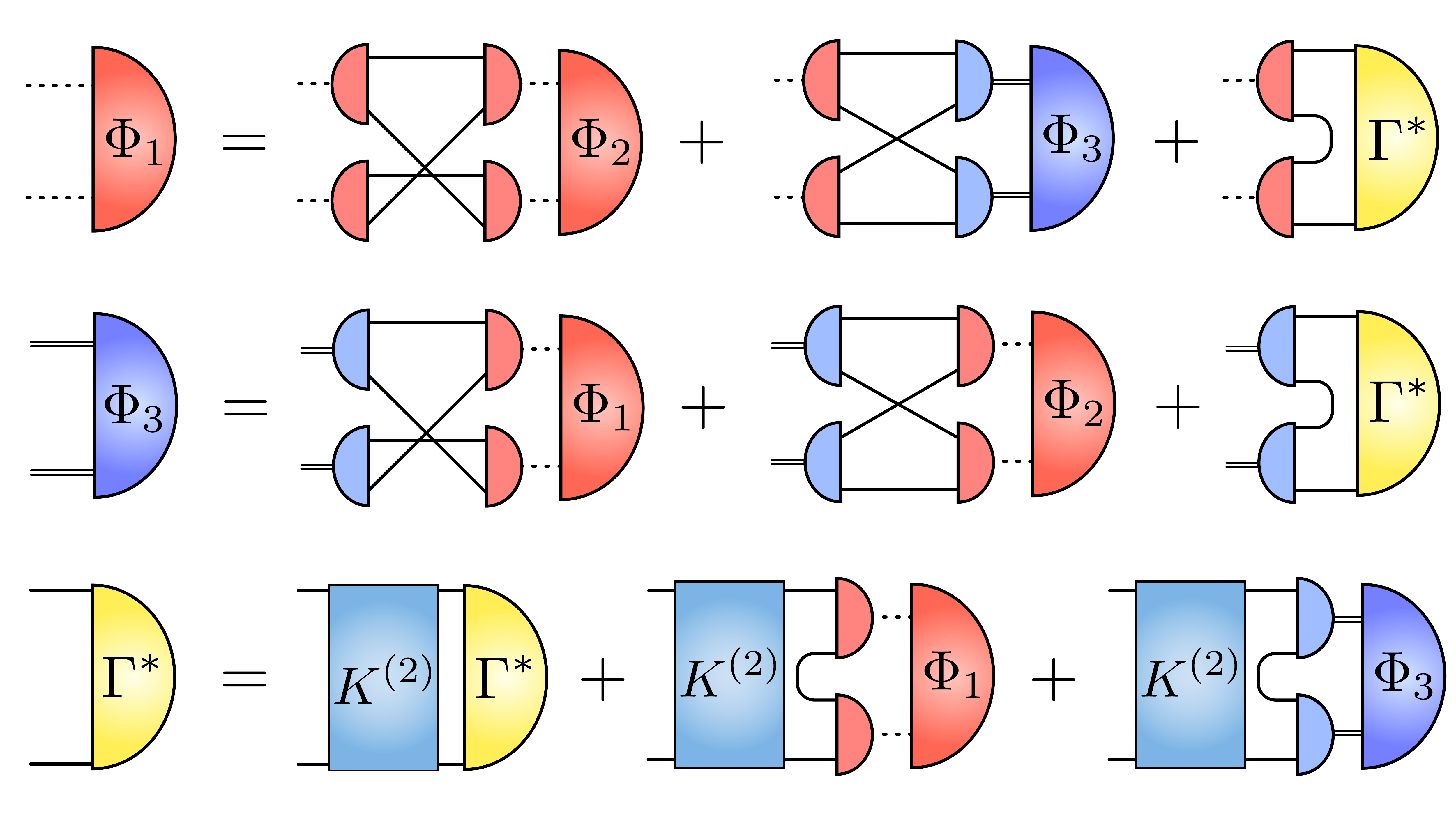}
	\caption{The coupled system of BSEs for one of the $\pi\pi$ contributions (red; first line) and the
		diquark/antidiquark contribution (blue; second line) to the four-quark state as well as the
		coupled BSE for the quark-antiquark state (yellow; third line) with the same quantum numbers.
		Not shown are the additional BSE for the second $\pi\pi$ contribution $\phi_2$ (which is redundant in our case)
		and another diagram for $\Gamma^*$ containing $\phi_2$ instead of $\phi_1$.   \label{fig-diagram2}}
\end{figure}

The corresponding two-quark $T$-matrix that contains this quarkonium state is determined by (cf. Eq.~(\ref{eq-dysonsum}) with $n=2$),
\begin{equation}
	T^{(2)}=\left(\mathds{1}-K^{(2)}G_0^{(2)}\right)^{-1}K^{(2)}\,.
\end{equation}
The corresponding four-quark $T$-matrix that contains the four-body component added in
Eq.~(\ref{eq-4bsastructure_v2}) is denoted by
\begin{equation}
T_a^{(4,2)} = K^{(2)}_{13} K^{(2)}_{24} S_{34} G_0^{(2)} T_{12}^{(2)} G_0^{(2)} S_{34} K^{(2)}_{24} K^{(2)}_{13}
\end{equation}
As a result, the master equation (\ref{eq-masterbse}) then contains the two-body equation for $\Gamma^*$ as an additional element and the equations
for the four-body meson-meson and diquark-antidiquark components of the full BSA are modified by additional terms containing $\Gamma^*$. The resulting
system of equations is shown diagrammatically in Fig.~\ref{fig-diagram2}. In its derivation we have  frequently used the BSE
\begin{equation}
\Gamma^{(2)} = K^{(2)}G_0^{(2)}\Gamma^{(2)}
\end{equation}
for the internal meson and diquark states in order to absorb two-body interaction kernels in the corresponding Bethe-Salpeter amplitudes.
We observe $q\bar{q}$ contributions in both BSEs for the meson-meson
 and  diquark-antidiquark components of the four-body amplitudes as well as a back-coupling of the meson-meson and diquark-antidiquark components into the
$q\bar{q}$ equation.

We wish to emphasise that at the present state the structure of the coupling between the four-quark state and the two-quark state 
is introduced by hand. There
is some formal motivation from the fact that the Dirac structure of the new elements in Eq.~(\ref{eq-4bsastructure_v2}) is part of the full
basis of the four-quark amplitude and thus constitutes an extension of the physically motivated basis of meson and diquark elements.
Formally, however, one would probably need to derive the putative coupling of four-quark and two-quark BSEs from a framework such as the
one given in Ref.~\cite{Yokojima:1993np}. This will be addressed in future work. Furthermore, it is interesting to observe that we would
have ended up with equations of similar structure if we had used expressions such as $\Gamma^*_{12} \,S_{34}^{-1}$ for the extension of
Eq.~(\ref{eq-4bsastructure_v2}) much along the line of $^3P_0$-models used frequently in the description of meson and baryon
decays \cite{Capstick:2000qj}. Such disconnected contributions have been considered already in Ref.~\cite{Kvinikhidze:2014yqa}
in the context of the original four-body equation. A potential problem with disconnected contributions is that they heavily restrict
the momentum structure of the four-body amplitude; this is avoided by our choice, Eq.~(\ref{eq-4bsastructure_v2}).

\section{\label{sec:3} Technical details}

\subsection{\label{sec:3.1} Quark propagator, mesons and diquarks}

The elements necessary to solve the coupled system of BSEs in Fig.~\ref{fig-diagram2} are dressed quark propagators,
meson and diquark Bethe-Salpeter amplitudes and
the corresponding propagators. All these ingredients are determined consistently from an underlying quark-gluon interaction that respects chiral
symmetry. The technical details of these types of calculations have been described in many works, see e.g. \cite{Maris:2003vk,Eichmann:2016yit,Sanchis-Alepuz:2017jjd} for reviews, thus we only give a short summary here.

We begin by specifying the Dyson-Schwinger equation for the dressed quark propagator,
\begin{equation}\label{quarkdse}
S^{-1}_{\alpha\beta}(p) = Z_2 \left( i\pslash + m_0 \right)_{\alpha\beta}
+ C_F \int_q \mathcal{K}_{\alpha\alpha'\beta'\beta} \,S_{\alpha'\beta'}(q)\,,
\end{equation}
with wave-function renormalization constant $Z_2$, bare quark mass $m_0$ and the Casimir $C_F=4/3$ for $N_c=3$ from the color trace.
In this form the equation is still exact
and the interaction kernel $\mathcal{K}_{\alpha\alpha'\beta'\beta}$ contains the dressed gluon propagator as well as
one bare and one dressed quark-gluon vertex. The Greek subscripts refer to color, flavor and Dirac structure. In previous
treatments of the four-quark problem \cite{Heupel:2012ua,Eichmann:2015cra,Wallbott:2019dng,Wallbott:2020jzh}, the rainbow-ladder
approximation has been used and we adopt the same interaction here. Then the kernel can be written as
\begin{equation}\label{RLkernel}
\mathcal{K}_{\alpha\alpha'\beta\beta'} =  Z_2^2 \, \frac{ 4\pi \alpha(k^2)}{k^2} \,
T^{\mu\nu}_k \gamma^\mu_{\alpha\alpha'} \,\gamma^\nu_{\beta\beta'},
\end{equation}
with the transverse projector $T^{\mu\nu}_k=\delta^{\mu\nu} - k^\mu k^\nu/k^2$. In this formulation, both the gluon dressing
function and the vector part $\sim \gamma^\mu$ of the quark-gluon vertex have been absorbed into an effective running
coupling $\alpha(k^2)$ which is taken from Ref.~\cite{Maris:1999nt} and has been discussed in detail e.g. in \cite{Eichmann:2016yit}.
This truncation guarantees the correct logarithmic behaviour of the quark at large momenta. Most importantly in the present context,
however, is that it allows for the preservation of the axialvector Ward-Takahashi identity by using the same interaction kernel
in the Bethe-Salpeter equations for the mesons and diquarks.

With the quark propagator from Eq.~(\ref{quarkdse}) and the quark-(anti-)quark interaction kernel Eq.~(\ref{RLkernel}) we then solve the
Bethe-Salpeter equations for light pseudoscalar mesons and scalar diquarks, which are the leading components (in terms of smallest masses)
of the two-body composition of our scalar four-quark state. The explicit representation of the BSA in terms of (four) Dirac, flavor and color
components as well as details on the technical treatment of meson BSEs can be found in the review articles \cite{Eichmann:2016yit,Sanchis-Alepuz:2017jjd}.
The meson/diquark propagators are then calculated via $T=\Gamma D \bar{\Gamma}$ and Eq.~(\ref{eq-dysonsum}). We obtain the following masses,
\begin{equation*}
m_\pi=0.138(2)\:\text{GeV}\qquad\qquad m_{\text{dq},0^+}=0.801(31)\:\text{GeV}\,,
\end{equation*}
and the corresponding BSAs. We are working in the isospin symmetric limit $m_u=m_d$.
Since the pseudoscalar meson and scalar diquark states are dominated by their leading Dirac structure, we only take those
into account in BSAs appearing internally in diagrams. For the external BSA in the two-body equation of the scalar meson
we use the full structure, i.e. all four amplitudes.

With this input, we are then in a position to solve the system of coupled integral equations for the BSEs in Fig.~\ref{fig-diagram2}.
Technically, this is done
as an eigenvalue problem, i.e. one introduces a general eigenvalue $\lambda$ on the left-hand side of the BSE and searches for the total
momentum $P^2=-M^2$ that corresponds to $\lambda=1$. This is then the mass of the bound state/resonance in the coupled four-body/two-body system.
One problem that appears in this search is the appearance of singularities in the plane of complex total momentum due to the internal meson
and diquark propagators. Although the exact locations of these poles are known (from the solutions of the corresponding meson/diquark BSEs),
it remains a highly non-trivial problem to perform the integrations numerically. To avoid this problem we determine the eigenvalue curve $\lambda(M^2)$
in the singularity-free region $M<2m$, where $m$ is the mass of the lightest meson/diquark ingredient and $M$ is the four-quark bound state mass. We then
extrapolate the resulting curve further into the time-like momentum domain using rational functions. This procedure has been tested extensively
for cases where one can actually do the calculation and turned out to be very stable for states with small masses such as the $\sigma$ meson
studied in this work.

\subsection{\label{sec:3.2} Into the complex $P^2$ plane}\label{complex}

A quantitatively reliable procedure to extend the calculation of $\lambda(P^2)$ beyond the limitations caused by singularities in the complex momentum
plane has been discussed in Ref.~\cite{Williams:2018adr}. There, the analytic structure of the complex $P^2$ plane has been explored for the example
of the $\rho$ meson, which has a relatively small width and is therefore well suited for a first exploration. The extension of these ideas to the
four-quark BSE (or even to the coupled system of $q\bar{q}q\bar{q}-q\bar{q}$ components) is not straight-forward and requires more conceptual
work. For the purpose of this exploratory work we therefore restrict ourselves to the $q\bar{q}$ component of the scalar meson and its associated BSE.
The drawback of this restriction is that we cannot hope for quantitative results. As will become clear in section \ref{res:mixing}, the physical
state is dominated by its four-quark components and any restriction to the two-quark components alone will not lead to masses that should be
compared with experiment. On the other hand, the technically much simpler BSE for the $q\bar{q}$ component allows us to directly study the generation
of widths in a different channel than the previously studied $\rho$ meson with the potential to test the method further into the complex $P^2$ region of the
second Riemann sheet. As will become clear in section \ref{res:decay}, this is indeed the case and leads to qualitative insights.
\begin{figure}\centering
	\includegraphics[width=.5\textwidth]{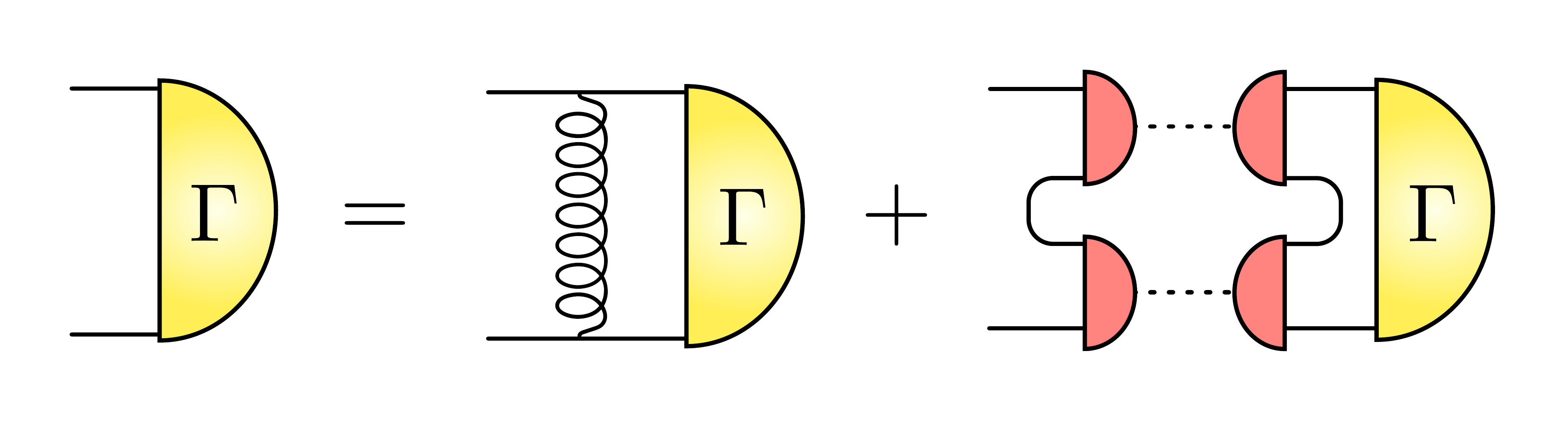}
	\caption{The BSE for the $q\bar{q}$ component of the scalar meson. Here, the pion decay term arises
		not from the coupling to the four-quark equation but from contributions beyond rainbow-ladder.\label{fig_brl_bse}}
\end{figure}

We begin by a short description of the BSE for the $q\bar{q}$ component of the scalar meson that we study. Its diagramatical representation is
shown in Fig.~\ref{fig_brl_bse}. In addition to the term with effective one-gluon exchange, which arises from the rainbow-ladder construction associated
with the kernel Eq.~(\ref{RLkernel}), we take into account a two-pion contribution that stems from corrections beyond rainbow-ladder. Its origin
is elucidated in detail in Refs.~\cite{Watson:2004kd,Watson:2004jq,Williams:2018adr,Miramontes:2019mco}.
Note that this term is different from the corresponding one
shown in Fig.~\ref{fig-diagram2} arising from the mixing with the four-body equation: whereas here the two-pion kernel couples directly to the
$q\bar{q}$ amplitude, in the latter approach it couples to the four-quark amplitude. However, both terms share a crucial property: they contain
two pions in the s-channel, which may go on-shell for certain loop momenta. As a consequence, we expect that both the mixed approach Fig.~\ref{fig-diagram2}
and the $q\bar{q}$ approach via Fig.~\ref{fig_brl_bse} lead to a similar structure in their respective complex plane of total momenta, namely
a cut at real timelike momenta starting from $2 m_\pi$ and a resonance state in the second Riemann sheet. Whereas in the mixed approach we can
not (yet) trace that resonance and can only extrapolate onto its real part (see the results of section \ref{res:mixing}), in the two-quark approach
we have this possibility (with results discussed in section \ref{res:decay}). Due to the similarities in structure of both approaches we therefore
expect qualitatively meaningful results from the $q\bar{q}$ approach that will carry over to the four-quark approach and can be used as guidance
for future work.

\begin{figure} \centering
	\includegraphics[width=0.7\columnwidth]{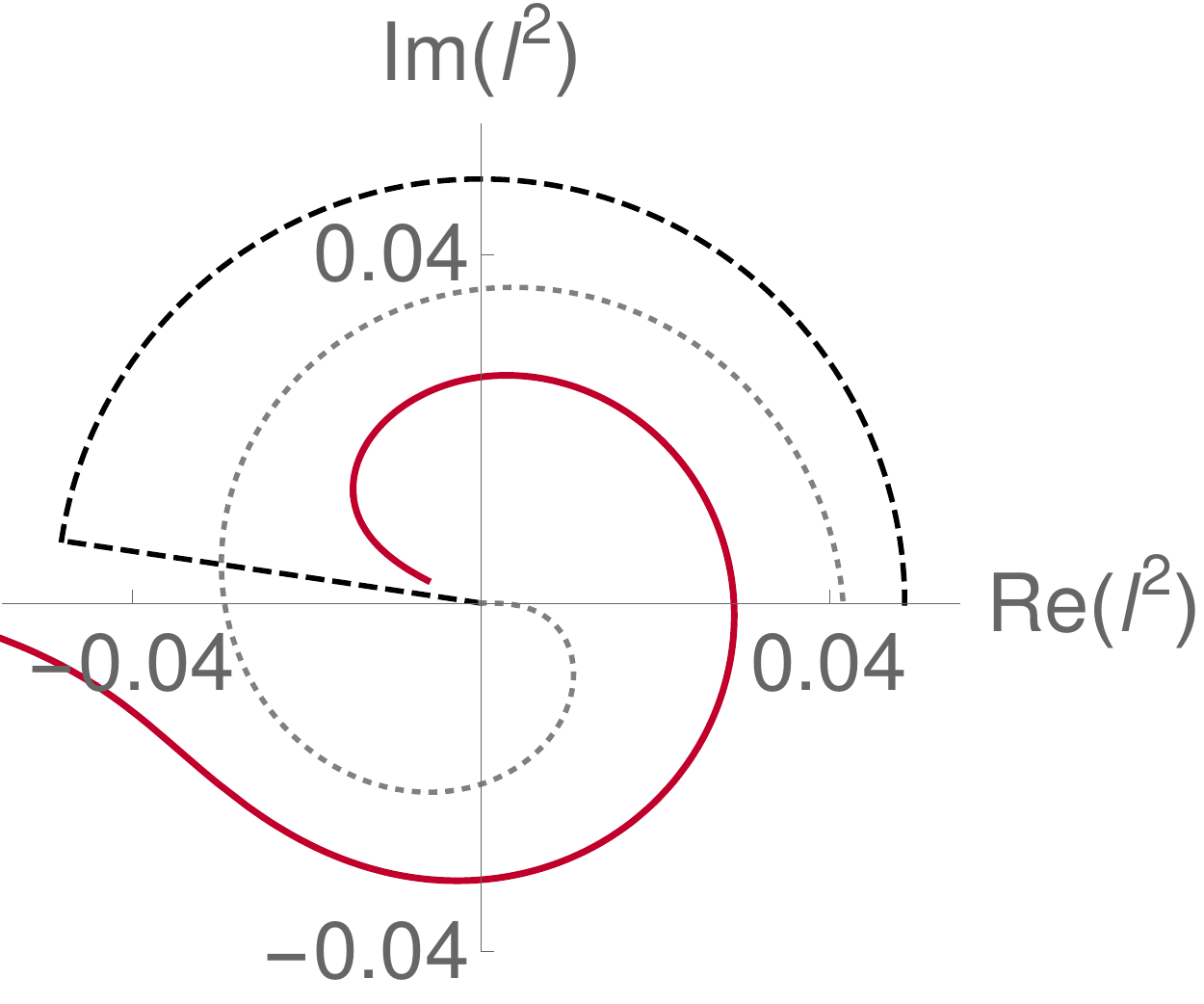}
	\caption{We show the two-pion branch cut (red solid line) and two possible paths (dashed) avoiding the cut in the complex $l^2$ plane.
		\label{fig_gamma_small_big}}
\end{figure}

We proceed by a description of the tools necessary to identify the decay width from the BSE of Fig.~\ref{fig_brl_bse}. The two-pion diagram
is given by
\begin{align}
\Gamma(P,p) =& \left(\dots \right) + C_{\pi\pi} F_{\pi\pi}  \int_l \left[\Gamma_1(l_+,2p-l_-) S(p-l) \right.\nonumber\\
            & \hspace*{2cm} \times \left. \bar{\Gamma}_2(-l_-,2p-l_+) \right] D(l_+) D(l_-) \nonumber \\
            & \times \int_q \left[ \bar{\Gamma}_1(-l_+,2q-l_-) S(q_+) \right.\nonumber\\
            & \hspace*{0cm} \times \left. \Gamma(P,q) S(q_-) \Gamma_2(l_-,2q-l_+) S(q-l) \right]
\label{eq_decay_diag}
\end{align}
with flavor and color projections already performed and the ellipsis denotes the rainbow-ladder diagram. We use the abbreviations
$\int_l = \int \frac{d^4 l}{(2\pi)^4}$, $l_{\pm} = l \pm P/2$, $q_{\pm} = q \pm P/2$ and $S$ denotes the quark propagator whereas
$D$ stands for the pion propagator,
\begin{align*}
D(p) = \left(p^2 + m_\pi^2 \right)^{-1}\,.
\end{align*}
In order for both integrations over $q$ and $l$ to avoid the pole locations of the pion propagators we need to analytically deform
their integration paths into the complex $l^2$ and $q^2$ planes respectively. After angular integration these poles generate
a branch cut that is shown in the $l^2$ plane in Fig.~\ref{fig_gamma_small_big} together with two possible choices of paths around
the cuts.
We have tested both choices and found them numerically equivalent.
\begin{figure}
	\includegraphics[width=\columnwidth]{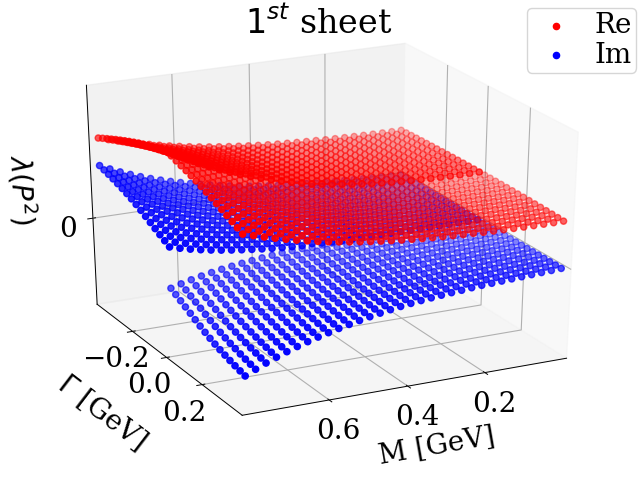}
	\caption{Eigenvalue curves $\lambda((M + i \Gamma/2)^2)$ of the BSE for the $\rho$ meson (and qualitatively similar for the $q\bar{q}$
		component of the $\sigma$-meson)
		in the first Riemann sheet. On top is the surface
		for the real part of $\lambda$, below the one for the imaginary part, where a branch cut opens at the two-pion threshold. \label{fig_rho}}
\end{figure}
\begin{figure*}[t]
	\centering
	\includegraphics[scale=1.3]{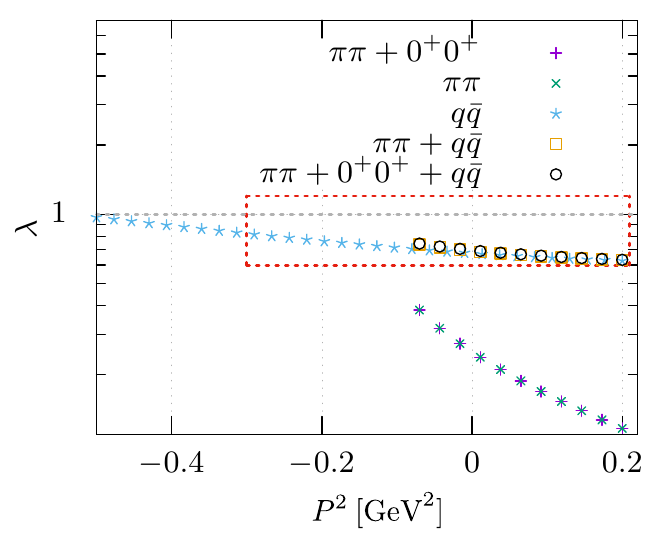}\hfill
	\includegraphics[scale=1.3]{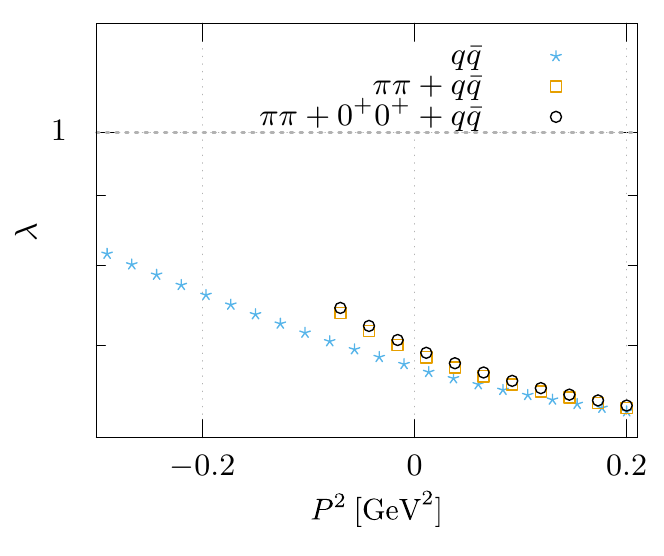}
	\caption{The eigenvalue curves $\lambda(P^2)$ for different combinations of mixing ingredients. Bound states/resonances occur at $P^2=-M^2$.
		Shown are results of calculations with different active components in the Bethe-Salpeter wave functions, see main text for details.
		The diagram on the right shows the same results as the diagram on the left but with the red area magnified.
		\label{fig-evcurves}}
\end{figure*}

With this path deformation we are now in a position to determine the eigenvalue curve $\lambda(P^2)$ on the first Riemann sheet. Using
the notation $P^2=(M + i \Gamma/2)^2$, we display the corresponding result for the $\rho$ meson determined in Ref.~\cite{Williams:2018adr}
in Fig.~\ref{fig_rho}. One clearly observes the branch cut opening at the two-pion threshold in the imaginary part of $\lambda$ (lower plane).

Unfortunately, from numerical solutions of the four-body BSE we cannot go beyond the branch cut and enter the second Riemann sheet 
(see \ref{res:decay} for a detailed explanation why). Instead, this is done
using a method of analytic continuation based on work by Thiele and Schlessinger
and employed recently in a number of publications, see e.g.~\cite{Tripolt:2016cya,Tripolt:2018xeo,Binosi:2019ecz,Eichmann:2019dts},
where details on the method can be found. In short, it amounts to using interpolating rational functions $R(x) = P(x)/Q(x)$ of degree $N$ to $N$
selected points of the function in question, thereby allowing for an analytical continuation beyond the region where the $N$ points
have been taken from. One of the problems that one encounters with this method is the appearance of `fake' singularities due to imprecise
cancellations in $P(x)$ and $Q(x)$. Since the location of these fake singularities and in addition also the corresponding residues
(which are typically small) depend heavily on the selection of points used in the fitting procedure, they can be identified and eliminated
by performing the procedure many times with random selections of points in a given region. What remains are potential `true' singularities.
The variation of the location of these stable singularities within the series of fitting procedures is quite stable and allows for a
statistical error estimate. We will discuss this further in the context of our results presented in section \ref{sec:4.2}.

\section{\label{sec:4} Results}

\subsection{\label{sec:4.1} The $q\bar{q}q\bar{q}-q\bar{q}$ mixing}\label{res:mixing}

Let us first come back to the mixed system of four- and two-quark states, Fig.~\ref{fig-diagram2}.
We have performed calculations for a $J^{PC}=0^{++}$ state consisting of light $u$ and $d$ quarks. As internal substructures of the four-quark state we
take into account two pions for the mesonic component and two scalar diquarks for the diquark component. Thus in total, our wave function contains
three components. In the practical calculation we can switch these on and off, such that we are able to gauge the impact of each individual component.
In Ref.~\cite{Heupel:2012ua} it has already been noted that the diquark component plays only a negligible role as compared to the two-pion contribution.
Here we need to study whether this is still the case once the mixing with the $q\bar{q}$ state is taken into account.

Our results for the corresponding
eigenvalue curves are displayed in the two diagrams of Fig.~\ref{fig-evcurves}. We first reproduced the results of Ref.~\cite{Heupel:2012ua}: the eigenvalue
curves for the two-pion contribution (green x) and the `two-pion plus diquark' case (violet cross) are on top of each other. This result changes slightly
in the full calculation using all three components, as can be seen by comparing the curves with two-pion plus $q\bar{q}$ (orange boxes) with the full result
(black circles). As can be seen in particular in the magnified diagram on the right, the two curves are no longer identical but small changes are indeed
induced by the diquark component. In terms of the (extrapolated) real parts of the masses of the resonance, this becomes evident in Table \ref{tab-comparison}.
Whereas the masses are identical (within error) in the two setups with four-quark contributions only, $M=416(26)\:$MeV\footnote{The small difference
of this result with the one reported in Ref.~\cite{Heupel:2012ua} is due to improved numerics and extrapolation procedures}, there are small
corrections of the order of 3 percent once the mixing with the quarkonium component is taken into account: $M=472\pm22$ vs. $M=456\pm24\:$MeV.
Thus, the overlap of the diquark components with the $q\bar{q}$ components (and therefore their indirect effect on the mass) is larger than 
their overlap with the $\pi\pi$ components. We do not have a deep explanation why this is the case. Anyway, their contributions 
remain very small.

It is also interesting to note that the eigenvalue curve of the full system follows the corresponding one for the quarkonium system (blue stars) for
spacelike values of $P^2=-M^2$ and only deviates shortly before we cannot follow the curve any more. By comparison with the curve without quarkonium
(violet crosses), it seems clear that this is the region where the two-pion component becomes important and begins to dominate the curve. This ties
in with fact that this is precisely the region where the pion poles enter the integration region of the system of BSEs. By comparing the results
of the extrapolation ($456\pm24$ vs. $416\pm26$) we find that although the quarkonium contribution is subdominant, it shifts the resulting
mass of the resonance by almost ten percent, which is quantitatively non-negligible.
\begin{table}[b]
	\begin{tabular}{|c|c|c|}
		\hline
		\textbf{[MeV]}				&\textbf{ground state mass} & \textbf{first excitation}\\\hline\hline
		$\pi\pi$					&	$416\pm26$				& $970\pm130$	\\\hline
		$\pi\pi+0^+0^+$				&	$416\pm26$				& $970\pm130$	\\\hline
		$q\bar{q}$					&	$667\pm2 $				& $1036\pm8$	\\\hline
		$\pi\pi+q\bar{q}$			&	$472\pm22$				& $1080\pm280$	\\\hline
		$\pi\pi+0^+0^++q\bar{q}$	&	$456\pm24$				& $1110\pm110$	\\\hline
	\end{tabular}
	\caption{The masses of different setups isolating and mixing four-quark components and quarkonia from and with each other.
		The error estimates stem from variations of the curve extrapolations using different rational functions. \label{tab-comparison}}
\end{table}

Nevertheless, the main conclusion of Ref.~\cite{Heupel:2012ua} remains intact and is in fact reaffirmed: the physical state is dominated by $\pi\pi$
components. Thus it must have a large decay width into two pions. Furthermore, its mass
\begin{equation}
M_{0^{++}} = 456\,\pm\,24 \,\mbox{MeV}
\end{equation}
is in the right ballpark to make the identification with
the $f_0(500)$ extracted from pion scattering experiments using dispersion theory at
$M=(449^{+22}_{-16}) - i\,(275 \pm 12)$ MeV \cite{Pelaez:2015qba} very plausible.
\begin{figure*}
	\includegraphics[width=\columnwidth]{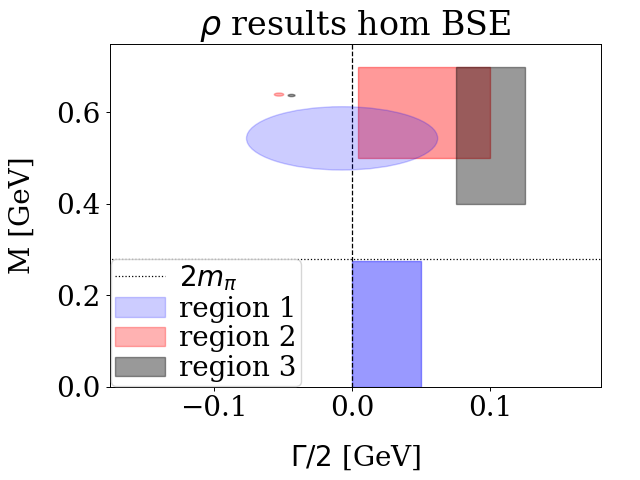}
	\includegraphics[width=\columnwidth]{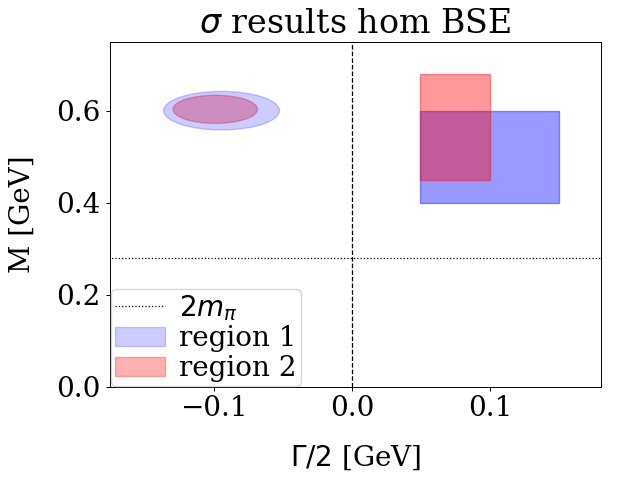}
	\caption{
		Location of the complex masses of the $\rho$ (left diagram) and the $\sigma$ meson (right diagram). Both plots show the $\Gamma<0$
		half plane of the second Riemann sheet on the left and the $\Gamma>0$ half plane in the first Riemann sheet on the right.
		Boxes on the right of each plot stand for data calculated from the BSE on the first Riemann sheet for $\Gamma>0$, circles
		on the left are the results obtained from analytic continuation to the second Riemann sheet for $\Gamma<0$.
		\label{fig_in_out}
	}
\end{figure*}

It is interesting to compare the results for the first radial excitation in all setups considered so far, given in the third column of
table \ref{tab-comparison}. In all combinations we do find a first radial excitation close to or even above one GeV. Most results have large
error bars due to the considerable distance of extrapolation involved from the region where we are able to determine the eigenvalues to the
region where they cross $\lambda=1$. This together with the fact that all excited states are close together makes a more detailed analysis
in analogy to the one discussed above impossible. Most importantly, we cannot say whether any component dominates the first excited state
of the full coupled system of BSEs. Nevertheless it is interesting to speculate what could happen if we augmented the rainbow-ladder
two-body kernel $K^{(2)}$ with corrections beyond rainbow-ladder. It is well known from BSE calculations of the $q\bar{q}$ system that the
pion, protected by chiral symmetry, does not receive any corrections beyond rainbow-ladder. On the other hand, the scalar pure $q\bar{q}$ 
state receives large additive corrections beyond rainbow-ladder of the order of 400 MeV \cite{Chang:2009zb,Fischer:2009jm,Williams:2015cvx}.
It is therefore plausible, that in the coupled system the ground state dominated by the $\pi\pi$ component will not change by much beyond
rainbow ladder. On the other hand, the first radial excitation dominated by the $q\bar{q}$ component will be substantially affected and thus
may end up in the region of $1400$ MeV. Thus the identification of our first radial excitation with the physical state $f_0(1370)$ seems
plausible.

\subsection{\label{sec:4.2} The decay width of the scalar}\label{res:decay}

As discussed in section \ref{complex}, we are not yet in a position to extract a width from the coupled system of equations Fig.~\ref{fig-diagram2} 
but need to resort the simpler $\bar{q}{q}$ BSE for the scalar state displayed in Fig.~\ref{fig_brl_bse}.
Our results for the resonance location from this scalar-meson BSE are shown in the right diagram of Fig.~\ref{fig_in_out}, whereas in the
left diagram we give corresponding results for the vector meson for comparison. The boxes mark the regions where data for the analytic
continuation were taken from,
which generate as output the corresponding locations of the resonance singularities marked by the ellipses. 

Let us first concentrate on the left
diagram. We clearly see that the error is very large when data are taken from region 1 below the two-pion threshold (indicated by the dotted line).
This large error bar makes it virtually impossible to obtain even qualitative results from the analytic
continuation in this case. This is the reason why we cannot apply the analytic
continuation procedure to the coupled system of two- and four-quark contributions discussed in the previous section, since for that system we only have data
available below the threshold. 
For the two-quark approach of section \ref{sec:3.2}, however, this is possible due to the path deformation described above.
We are then able to access the regions 2 and 3, which lead to the very small ellipses shown in the plot. Our results,
\begin{align*}
M_{\rho} = 638(2) \  \text{MeV}\,, \qquad \Gamma_{\rho} = 108 (4) \ \text{MeV}\,,
\end{align*}
are in good agreement with the one obtained in \cite{Williams:2018adr} using similar methods. Compared to the experimental values, our mass is
too small by about 130 MeV and the width by about 40 MeV. This does not come completely as a surprise since it is known that beyond rainbow-ladder
corrections due to the gluon self-interactions add significantly to the mass of the vector \cite{Fischer:2009jm}. Whether further corrections
are also able to account for the missing width remains to be studied in the future.

The width extracted for the scalar $\sigma$ meson is displayed in the right diagram of Fig.~\ref{fig_in_out}.
Again, the shaded region on the right is the selected region where we calculate data from the BSE on the
first Riemann sheet. Ellipses  on the left stand for the results in the second sheet obtained from our analytic
continuation procedure, where errors are proportional to the size of the ellipses. Comparing with the vector case,
we observe a significantly larger width. Correspondingly, the errors are also larger. This is to be expected, since
the extrapolation needs to probe further into the second Riemann sheet. The results are fairly stable with respect to different
choices of input regions, however regions below threshold do not work well at all, similar to the vector case.
The final result is obtained by performing 30 runs, each one with a fixed number of input points within the region
$M\in [0.45,0.68]$ GeV and $\Gamma \in [0.1,0.2]$ GeV. We do this 10 times and obtain a $\sigma$ mass and an error
in the form of a standard deviation each time. We then take the average of the ten resulting $\sigma$ masses and calculate
the final error using standard error propagation. We obtain:
\begin{align*}
M_{\sigma} = 587(12) \  \text{MeV}, \qquad \Gamma_{\sigma} = 186 (24) \ \text{MeV}.
\end{align*}

It is interesting to compare our results against the calculations from $\pi\pi$ scattering in rainbow-ladder from \cite{Cotanch:2002vj}.
The (real part of the) masses are (almost) identical to our results. Furthermore, a width was calculated in \cite{Cotanch:2002vj,Mader:2011zf}
in the impulse approximation, i.e. using pure rainbow-ladder Bethe-Salpeter amplitudes inside triangle decay diagrams. The resulting
widths are similar to the ones we extracted from the second Riemann sheet.

As already discussed above, our result is rather qualitative than quantitative. The real part of the mass is too large, while the width
is too small by more than a factor of two. Of course, these two findings may very well be related to each other and we therefore may
hope that solving the full coupled system of equations, Fig.~\ref{fig-diagram2}, in the complex $P^2$ plane may yield much more
quantitative results for the width. Since the numerical effort involved is considerable, this is left for future work.

\section{\label{sec:5} Summary}\label{sec:sum}
In this work we have addressed the mass and the width of the lowest-lying scalar meson in a framework that uses functional methods,
i.e. the Dyson-Schwinger and Bethe-Salpeter equations of QCD. We have solved a coupled system of BSEs that combines contributions
from four-quark components (in the form of $\pi\pi$ and diquark-antidiquark amplitudes) and quark-antiquark components. We emphasize
that this counting of components is not a counting with respect to Fock states, but that both the four-quark and two-quark
contributions each summarize contributions from infinitely many quarks, antiquarks and gluons. What distinguishes the four-
and two-quark contributions instead is the associated valence-quark content needed to build the quantum numbers of the scalar.
Our results confirm previous findings from Ref.~\cite{Heupel:2012ua}. The physical state is dominated by the $\pi\pi$ component
and the diquark-antidiquark component is almost negligible. Subleading contributions of the order of ten percent of the total mass
stem from the quark-antiquark component of the Bethe-Salpeter wave function. This finding explains the small mass of the
state, since it is dominated by pseudo-Goldstone bosons, and its large decay width into two pions. We have confirmed the latter
property (in a simplified system) explicitly by analytically continuing to the second Riemann sheet and determining the associated singularity
in the complex energy plane. Our results thus suggest an identification with the experimental $\sigma$ meson, the $f_0(500)$.
Moreover, we found a first radial excitation in the coupled system of BSEs which may be identified with the $f_0(1370)$, although
with much larger uncertainty than is the case for the $\sigma$ meson.
\subsection*{Acknowledgements}
We are grateful to Sasa Prelovsek, George Rupp and Marc Wagner for discussions. We are also grateful to the organizers
of the very interesting conference `Excited QCD 2020, February 2-8 Krynica Zdroj, Poland', where the idea for part of this work has been born.
This work was supported by the DFG grant FI 970/11-1, by the Helmholtz International Center for FAIR within
the LOEWE program of the State of Hesse, by the Helmholtz Research Academy Hesse for FAIR (HFHF), and by the FCT Investigator Grant IF/00898/2015.

\bibliographystyle{apsrev4-1}
\bibliography{sigma_mixing}

\begin{thebibliography}{50}%
\makeatletter
\providecommand \@ifxundefined [1]{%
 \@ifx{#1\undefined}
}%
\providecommand \@ifnum [1]{%
 \ifnum #1\expandafter \@firstoftwo
 \else \expandafter \@secondoftwo
 \fi
}%
\providecommand \@ifx [1]{%
 \ifx #1\expandafter \@firstoftwo
 \else \expandafter \@secondoftwo
 \fi
}%
\providecommand \natexlab [1]{#1}%
\providecommand \enquote  [1]{``#1''}%
\providecommand \bibnamefont  [1]{#1}%
\providecommand \bibfnamefont [1]{#1}%
\providecommand \citenamefont [1]{#1}%
\providecommand \href@noop [0]{\@secondoftwo}%
\providecommand \href [0]{\begingroup \@sanitize@url \@href}%
\providecommand \@href[1]{\@@startlink{#1}\@@href}%
\providecommand \@@href[1]{\endgroup#1\@@endlink}%
\providecommand \@sanitize@url [0]{\catcode `\\12\catcode `\$12\catcode
  `\&12\catcode `\#12\catcode `\^12\catcode `\_12\catcode `\%12\relax}%
\providecommand \@@startlink[1]{}%
\providecommand \@@endlink[0]{}%
\providecommand \url  [0]{\begingroup\@sanitize@url \@url }%
\providecommand \@url [1]{\endgroup\@href {#1}{\urlprefix }}%
\providecommand \urlprefix  [0]{URL }%
\providecommand \Eprint [0]{\href }%
\providecommand \doibase [0]{http://dx.doi.org/}%
\providecommand \selectlanguage [0]{\@gobble}%
\providecommand \bibinfo  [0]{\@secondoftwo}%
\providecommand \bibfield  [0]{\@secondoftwo}%
\providecommand \translation [1]{[#1]}%
\providecommand \BibitemOpen [0]{}%
\providecommand \bibitemStop [0]{}%
\providecommand \bibitemNoStop [0]{.\EOS\space}%
\providecommand \EOS [0]{\spacefactor3000\relax}%
\providecommand \BibitemShut  [1]{\csname bibitem#1\endcsname}%
\let\auto@bib@innerbib\@empty
%</preamble>
\bibitem [{\citenamefont {Heupel}\ \emph {et~al.}(2012)\citenamefont {Heupel},
  \citenamefont {Eichmann},\ and\ \citenamefont {Fischer}}]{Heupel:2012ua}%
  \BibitemOpen
  \bibfield  {author} {\bibinfo {author} {\bibfnamefont {W.}~\bibnamefont
  {Heupel}}, \bibinfo {author} {\bibfnamefont {G.}~\bibnamefont {Eichmann}}, \
  and\ \bibinfo {author} {\bibfnamefont {C.~S.}\ \bibnamefont {Fischer}},\
  }\href {\doibase 10.1016/j.physletb.2012.11.009} {\bibfield  {journal}
  {\bibinfo  {journal} {Phys. Lett.}\ }\textbf {\bibinfo {volume} {B718}},\
  \bibinfo {pages} {545} (\bibinfo {year} {2012})},\ \Eprint
  {http://arxiv.org/abs/1206.5129} {arXiv:1206.5129 [hep-ph]} \BibitemShut
  {NoStop}%
%%CITATION = ARXIV:1206.5129;%%
\bibitem [{\citenamefont {Pelaez}(2016)}]{Pelaez:2015qba}%
  \BibitemOpen
  \bibfield  {author} {\bibinfo {author} {\bibfnamefont {J.}~\bibnamefont
  {Pelaez}},\ }\href {\doibase 10.1016/j.physrep.2016.09.001} {\bibfield
  {journal} {\bibinfo  {journal} {Phys. Rept.}\ }\textbf {\bibinfo {volume}
  {658}},\ \bibinfo {pages} {1} (\bibinfo {year} {2016})},\ \Eprint
  {http://arxiv.org/abs/1510.00653} {arXiv:1510.00653 [hep-ph]} \BibitemShut
  {NoStop}%
\bibitem [{\citenamefont {Caprini}\ \emph {et~al.}(2006)\citenamefont
  {Caprini}, \citenamefont {Colangelo},\ and\ \citenamefont
  {Leutwyler}}]{Caprini:2005zr}%
  \BibitemOpen
  \bibfield  {author} {\bibinfo {author} {\bibfnamefont {I.}~\bibnamefont
  {Caprini}}, \bibinfo {author} {\bibfnamefont {G.}~\bibnamefont {Colangelo}},
  \ and\ \bibinfo {author} {\bibfnamefont {H.}~\bibnamefont {Leutwyler}},\
  }\href {\doibase 10.1103/PhysRevLett.96.132001} {\bibfield  {journal}
  {\bibinfo  {journal} {Phys. Rev. Lett.}\ }\textbf {\bibinfo {volume} {96}},\
  \bibinfo {pages} {132001} (\bibinfo {year} {2006})},\ \Eprint
  {http://arxiv.org/abs/hep-ph/0512364} {arXiv:hep-ph/0512364} \BibitemShut
  {NoStop}%
\bibitem [{\citenamefont {Yndurain}\ \emph {et~al.}(2007)\citenamefont
  {Yndurain}, \citenamefont {Garcia-Martin},\ and\ \citenamefont
  {Pelaez}}]{Yndurain:2007qm}%
  \BibitemOpen
  \bibfield  {author} {\bibinfo {author} {\bibfnamefont {F.}~\bibnamefont
  {Yndurain}}, \bibinfo {author} {\bibfnamefont {R.}~\bibnamefont
  {Garcia-Martin}}, \ and\ \bibinfo {author} {\bibfnamefont {J.}~\bibnamefont
  {Pelaez}},\ }\href {\doibase 10.1103/PhysRevD.76.074034} {\bibfield
  {journal} {\bibinfo  {journal} {Phys. Rev. D}\ }\textbf {\bibinfo {volume}
  {76}},\ \bibinfo {pages} {074034} (\bibinfo {year} {2007})},\ \Eprint
  {http://arxiv.org/abs/hep-ph/0701025} {arXiv:hep-ph/0701025} \BibitemShut
  {NoStop}%
\bibitem [{\citenamefont {Garcia-Martin}\ \emph {et~al.}(2011)\citenamefont
  {Garcia-Martin}, \citenamefont {Kaminski}, \citenamefont {Pelaez},\ and\
  \citenamefont {Ruiz~de Elvira}}]{GarciaMartin:2011jx}%
  \BibitemOpen
  \bibfield  {author} {\bibinfo {author} {\bibfnamefont {R.}~\bibnamefont
  {Garcia-Martin}}, \bibinfo {author} {\bibfnamefont {R.}~\bibnamefont
  {Kaminski}}, \bibinfo {author} {\bibfnamefont {J.}~\bibnamefont {Pelaez}}, \
  and\ \bibinfo {author} {\bibfnamefont {J.}~\bibnamefont {Ruiz~de Elvira}},\
  }\href {\doibase 10.1103/PhysRevLett.107.072001} {\bibfield  {journal}
  {\bibinfo  {journal} {Phys. Rev. Lett.}\ }\textbf {\bibinfo {volume} {107}},\
  \bibinfo {pages} {072001} (\bibinfo {year} {2011})},\ \Eprint
  {http://arxiv.org/abs/1107.1635} {arXiv:1107.1635 [hep-ph]} \BibitemShut
  {NoStop}%
\bibitem [{\citenamefont {Moussallam}(2011)}]{Moussallam:2011zg}%
  \BibitemOpen
  \bibfield  {author} {\bibinfo {author} {\bibfnamefont {B.}~\bibnamefont
  {Moussallam}},\ }\href {\doibase 10.1140/epjc/s10052-011-1814-z} {\bibfield
  {journal} {\bibinfo  {journal} {Eur. Phys. J. C}\ }\textbf {\bibinfo {volume}
  {71}},\ \bibinfo {pages} {1814} (\bibinfo {year} {2011})},\ \Eprint
  {http://arxiv.org/abs/1110.6074} {arXiv:1110.6074 [hep-ph]} \BibitemShut
  {NoStop}%
\bibitem [{\citenamefont {Jaffe}(1977)}]{Jaffe:1976ig}%
  \BibitemOpen
  \bibfield  {author} {\bibinfo {author} {\bibfnamefont {R.~L.}\ \bibnamefont
  {Jaffe}},\ }\href {\doibase 10.1103/PhysRevD.15.267} {\bibfield  {journal}
  {\bibinfo  {journal} {Phys. Rev. D}\ }\textbf {\bibinfo {volume} {15}},\
  \bibinfo {pages} {267} (\bibinfo {year} {1977})}\BibitemShut {NoStop}%
\bibitem [{\citenamefont {Achasov}\ and\ \citenamefont
  {Ivanchenko}(1989)}]{Achasov:1987ts}%
  \BibitemOpen
  \bibfield  {author} {\bibinfo {author} {\bibfnamefont {N.}~\bibnamefont
  {Achasov}}\ and\ \bibinfo {author} {\bibfnamefont {V.}~\bibnamefont
  {Ivanchenko}},\ }\href {\doibase 10.1016/0550-3213(89)90364-7} {\bibfield
  {journal} {\bibinfo  {journal} {Nucl. Phys. B}\ }\textbf {\bibinfo {volume}
  {315}},\ \bibinfo {pages} {465} (\bibinfo {year} {1989})}\BibitemShut
  {NoStop}%
\bibitem [{\citenamefont {Black}\ \emph {et~al.}(1999)\citenamefont {Black},
  \citenamefont {Fariborz}, \citenamefont {Sannino},\ and\ \citenamefont
  {Schechter}}]{Black:1998wt}%
  \BibitemOpen
  \bibfield  {author} {\bibinfo {author} {\bibfnamefont {D.}~\bibnamefont
  {Black}}, \bibinfo {author} {\bibfnamefont {A.~H.}\ \bibnamefont {Fariborz}},
  \bibinfo {author} {\bibfnamefont {F.}~\bibnamefont {Sannino}}, \ and\
  \bibinfo {author} {\bibfnamefont {J.}~\bibnamefont {Schechter}},\ }\href
  {\doibase 10.1103/PhysRevD.59.074026} {\bibfield  {journal} {\bibinfo
  {journal} {Phys. Rev. D}\ }\textbf {\bibinfo {volume} {59}},\ \bibinfo
  {pages} {074026} (\bibinfo {year} {1999})},\ \Eprint
  {http://arxiv.org/abs/hep-ph/9808415} {arXiv:hep-ph/9808415} \BibitemShut
  {NoStop}%
\bibitem [{\citenamefont {Maiani}\ \emph {et~al.}(2004)\citenamefont {Maiani},
  \citenamefont {Piccinini}, \citenamefont {Polosa},\ and\ \citenamefont
  {Riquer}}]{Maiani:2004uc}%
  \BibitemOpen
  \bibfield  {author} {\bibinfo {author} {\bibfnamefont {L.}~\bibnamefont
  {Maiani}}, \bibinfo {author} {\bibfnamefont {F.}~\bibnamefont {Piccinini}},
  \bibinfo {author} {\bibfnamefont {A.}~\bibnamefont {Polosa}}, \ and\ \bibinfo
  {author} {\bibfnamefont {V.}~\bibnamefont {Riquer}},\ }\href {\doibase
  10.1103/PhysRevLett.93.212002} {\bibfield  {journal} {\bibinfo  {journal}
  {Phys. Rev. Lett.}\ }\textbf {\bibinfo {volume} {93}},\ \bibinfo {pages}
  {212002} (\bibinfo {year} {2004})},\ \Eprint
  {http://arxiv.org/abs/hep-ph/0407017} {arXiv:hep-ph/0407017} \BibitemShut
  {NoStop}%
\bibitem [{\citenamefont {Giacosa}(2006)}]{Giacosa:2006rg}%
  \BibitemOpen
  \bibfield  {author} {\bibinfo {author} {\bibfnamefont {F.}~\bibnamefont
  {Giacosa}},\ }\href {\doibase 10.1103/PhysRevD.74.014028} {\bibfield
  {journal} {\bibinfo  {journal} {Phys. Rev. D}\ }\textbf {\bibinfo {volume}
  {74}},\ \bibinfo {pages} {014028} (\bibinfo {year} {2006})},\ \Eprint
  {http://arxiv.org/abs/hep-ph/0605191} {arXiv:hep-ph/0605191} \BibitemShut
  {NoStop}%
\bibitem [{\citenamefont {Klempt}\ and\ \citenamefont
  {Zaitsev}(2007)}]{Klempt:2007cp}%
  \BibitemOpen
  \bibfield  {author} {\bibinfo {author} {\bibfnamefont {E.}~\bibnamefont
  {Klempt}}\ and\ \bibinfo {author} {\bibfnamefont {A.}~\bibnamefont
  {Zaitsev}},\ }\href {\doibase 10.1016/j.physrep.2007.07.006} {\bibfield
  {journal} {\bibinfo  {journal} {Phys. Rept.}\ }\textbf {\bibinfo {volume}
  {454}},\ \bibinfo {pages} {1} (\bibinfo {year} {2007})},\ \Eprint
  {http://arxiv.org/abs/0708.4016} {arXiv:0708.4016 [hep-ph]} \BibitemShut
  {NoStop}%
\bibitem [{\citenamefont {Ebert}\ \emph {et~al.}(2009)\citenamefont {Ebert},
  \citenamefont {Faustov},\ and\ \citenamefont {Galkin}}]{Ebert:2008id}%
  \BibitemOpen
  \bibfield  {author} {\bibinfo {author} {\bibfnamefont {D.}~\bibnamefont
  {Ebert}}, \bibinfo {author} {\bibfnamefont {R.}~\bibnamefont {Faustov}}, \
  and\ \bibinfo {author} {\bibfnamefont {V.}~\bibnamefont {Galkin}},\ }\href
  {\doibase 10.1140/epjc/s10052-009-0925-2} {\bibfield  {journal} {\bibinfo
  {journal} {Eur. Phys. J. C}\ }\textbf {\bibinfo {volume} {60}},\ \bibinfo
  {pages} {273} (\bibinfo {year} {2009})},\ \Eprint
  {http://arxiv.org/abs/0812.2116} {arXiv:0812.2116 [hep-ph]} \BibitemShut
  {NoStop}%
\bibitem [{\citenamefont {Alford}\ and\ \citenamefont
  {Jaffe}(2003)}]{Alford:2003xw}%
  \BibitemOpen
  \bibfield  {author} {\bibinfo {author} {\bibfnamefont {M.}~\bibnamefont
  {Alford}}\ and\ \bibinfo {author} {\bibfnamefont {R.}~\bibnamefont {Jaffe}},\
  }\href {\doibase 10.1063/1.1632207} {\bibfield  {journal} {\bibinfo
  {journal} {AIP Conf. Proc.}\ }\textbf {\bibinfo {volume} {688}},\ \bibinfo
  {pages} {208} (\bibinfo {year} {2003})},\ \Eprint
  {http://arxiv.org/abs/hep-lat/0306037} {arXiv:hep-lat/0306037} \BibitemShut
  {NoStop}%
\bibitem [{\citenamefont {Mathur}\ \emph {et~al.}(2007)\citenamefont {Mathur},
  \citenamefont {Alexandru}, \citenamefont {Chen}, \citenamefont {Dong},
  \citenamefont {Draper}, \citenamefont {Horvath}, \citenamefont {Lee},
  \citenamefont {Liu}, \citenamefont {Tamhankar},\ and\ \citenamefont
  {Zhang}}]{Mathur:2006bs}%
  \BibitemOpen
  \bibfield  {author} {\bibinfo {author} {\bibfnamefont {N.}~\bibnamefont
  {Mathur}}, \bibinfo {author} {\bibfnamefont {A.}~\bibnamefont {Alexandru}},
  \bibinfo {author} {\bibfnamefont {Y.}~\bibnamefont {Chen}}, \bibinfo {author}
  {\bibfnamefont {S.}~\bibnamefont {Dong}}, \bibinfo {author} {\bibfnamefont
  {T.}~\bibnamefont {Draper}}, \bibinfo {author} {\bibfnamefont
  {I.}~\bibnamefont {Horvath}}, \bibinfo {author} {\bibfnamefont
  {F.}~\bibnamefont {Lee}}, \bibinfo {author} {\bibfnamefont {K.}~\bibnamefont
  {Liu}}, \bibinfo {author} {\bibfnamefont {S.}~\bibnamefont {Tamhankar}}, \
  and\ \bibinfo {author} {\bibfnamefont {J.}~\bibnamefont {Zhang}},\ }\href
  {\doibase 10.1103/PhysRevD.76.114505} {\bibfield  {journal} {\bibinfo
  {journal} {Phys. Rev. D}\ }\textbf {\bibinfo {volume} {76}},\ \bibinfo
  {pages} {114505} (\bibinfo {year} {2007})},\ \Eprint
  {http://arxiv.org/abs/hep-ph/0607110} {arXiv:hep-ph/0607110} \BibitemShut
  {NoStop}%
\bibitem [{\citenamefont {Prelovsek}(2010)}]{Prelovsek:2010ty}%
  \BibitemOpen
  \bibfield  {author} {\bibinfo {author} {\bibfnamefont {S.}~\bibnamefont
  {Prelovsek}},\ }\href@noop {} {\bibfield  {journal} {\bibinfo  {journal}
  {Acta Phys. Polon. Supp.}\ }\textbf {\bibinfo {volume} {3}},\ \bibinfo
  {pages} {975} (\bibinfo {year} {2010})},\ \Eprint
  {http://arxiv.org/abs/1004.3636} {arXiv:1004.3636 [hep-lat]} \BibitemShut
  {NoStop}%
\bibitem [{\citenamefont {Prelovsek}\ \emph {et~al.}(2010)\citenamefont
  {Prelovsek}, \citenamefont {Draper}, \citenamefont {Lang}, \citenamefont
  {Limmer}, \citenamefont {Liu}, \citenamefont {Mathur},\ and\ \citenamefont
  {Mohler}}]{Prelovsek:2010kg}%
  \BibitemOpen
  \bibfield  {author} {\bibinfo {author} {\bibfnamefont {S.}~\bibnamefont
  {Prelovsek}}, \bibinfo {author} {\bibfnamefont {T.}~\bibnamefont {Draper}},
  \bibinfo {author} {\bibfnamefont {C.~B.}\ \bibnamefont {Lang}}, \bibinfo
  {author} {\bibfnamefont {M.}~\bibnamefont {Limmer}}, \bibinfo {author}
  {\bibfnamefont {K.-F.}\ \bibnamefont {Liu}}, \bibinfo {author} {\bibfnamefont
  {N.}~\bibnamefont {Mathur}}, \ and\ \bibinfo {author} {\bibfnamefont
  {D.}~\bibnamefont {Mohler}},\ }\href {\doibase 10.1103/PhysRevD.82.094507}
  {\bibfield  {journal} {\bibinfo  {journal} {Phys. Rev.}\ }\textbf {\bibinfo
  {volume} {D82}},\ \bibinfo {pages} {094507} (\bibinfo {year} {2010})},\
  \Eprint {http://arxiv.org/abs/1005.0948} {arXiv:1005.0948 [hep-lat]}
  \BibitemShut {NoStop}%
%%CITATION = ARXIV:1005.0948;%%
\bibitem [{\citenamefont {Eichmann}\ \emph
  {et~al.}(2016{\natexlab{a}})\citenamefont {Eichmann}, \citenamefont
  {Fischer},\ and\ \citenamefont {Heupel}}]{Eichmann:2015cra}%
  \BibitemOpen
  \bibfield  {author} {\bibinfo {author} {\bibfnamefont {G.}~\bibnamefont
  {Eichmann}}, \bibinfo {author} {\bibfnamefont {C.~S.}\ \bibnamefont
  {Fischer}}, \ and\ \bibinfo {author} {\bibfnamefont {W.}~\bibnamefont
  {Heupel}},\ }\href {\doibase 10.1016/j.physletb.2015.12.036} {\bibfield
  {journal} {\bibinfo  {journal} {Phys. Lett.}\ }\textbf {\bibinfo {volume}
  {B753}},\ \bibinfo {pages} {282} (\bibinfo {year} {2016}{\natexlab{a}})},\
  \Eprint {http://arxiv.org/abs/1508.07178} {arXiv:1508.07178 [hep-ph]}
  \BibitemShut {NoStop}%
%%CITATION = ARXIV:1508.07178;%%
\bibitem [{\citenamefont {Close}\ and\ \citenamefont
  {Tornqvist}(2002)}]{Close:2002zu}%
  \BibitemOpen
  \bibfield  {author} {\bibinfo {author} {\bibfnamefont {F.~E.}\ \bibnamefont
  {Close}}\ and\ \bibinfo {author} {\bibfnamefont {N.~A.}\ \bibnamefont
  {Tornqvist}},\ }\href {\doibase 10.1088/0954-3899/28/10/201} {\bibfield
  {journal} {\bibinfo  {journal} {J. Phys. G}\ }\textbf {\bibinfo {volume}
  {28}},\ \bibinfo {pages} {R249} (\bibinfo {year} {2002})},\ \Eprint
  {http://arxiv.org/abs/hep-ph/0204205} {arXiv:hep-ph/0204205} \BibitemShut
  {NoStop}%
\bibitem [{\citenamefont {Giacosa}(2007)}]{Giacosa:2006tf}%
  \BibitemOpen
  \bibfield  {author} {\bibinfo {author} {\bibfnamefont {F.}~\bibnamefont
  {Giacosa}},\ }\href {\doibase 10.1103/PhysRevD.75.054007} {\bibfield
  {journal} {\bibinfo  {journal} {Phys. Rev. D}\ }\textbf {\bibinfo {volume}
  {75}},\ \bibinfo {pages} {054007} (\bibinfo {year} {2007})},\ \Eprint
  {http://arxiv.org/abs/hep-ph/0611388} {arXiv:hep-ph/0611388} \BibitemShut
  {NoStop}%
\bibitem [{\citenamefont {'t~Hooft}\ \emph {et~al.}(2008)\citenamefont
  {'t~Hooft}, \citenamefont {Isidori}, \citenamefont {Maiani}, \citenamefont
  {Polosa},\ and\ \citenamefont {Riquer}}]{Hooft:2008we}%
  \BibitemOpen
  \bibfield  {author} {\bibinfo {author} {\bibfnamefont {G.}~\bibnamefont
  {'t~Hooft}}, \bibinfo {author} {\bibfnamefont {G.}~\bibnamefont {Isidori}},
  \bibinfo {author} {\bibfnamefont {L.}~\bibnamefont {Maiani}}, \bibinfo
  {author} {\bibfnamefont {A.~D.}\ \bibnamefont {Polosa}}, \ and\ \bibinfo
  {author} {\bibfnamefont {V.}~\bibnamefont {Riquer}},\ }\href {\doibase
  10.1016/j.physletb.2008.03.036} {\bibfield  {journal} {\bibinfo  {journal}
  {Phys. Lett.}\ }\textbf {\bibinfo {volume} {B662}},\ \bibinfo {pages} {424}
  (\bibinfo {year} {2008})},\ \Eprint {http://arxiv.org/abs/0801.2288}
  {arXiv:0801.2288 [hep-ph]} \BibitemShut {NoStop}%
%%CITATION = ARXIV:0801.2288;%%
\bibitem [{\citenamefont {Londergan}\ \emph {et~al.}(2014)\citenamefont
  {Londergan}, \citenamefont {Nebreda}, \citenamefont {Pelaez},\ and\
  \citenamefont {Szczepaniak}}]{Londergan:2013dza}%
  \BibitemOpen
  \bibfield  {author} {\bibinfo {author} {\bibfnamefont {J.}~\bibnamefont
  {Londergan}}, \bibinfo {author} {\bibfnamefont {J.}~\bibnamefont {Nebreda}},
  \bibinfo {author} {\bibfnamefont {J.}~\bibnamefont {Pelaez}}, \ and\ \bibinfo
  {author} {\bibfnamefont {A.}~\bibnamefont {Szczepaniak}},\ }\href {\doibase
  10.1016/j.physletb.2013.12.061} {\bibfield  {journal} {\bibinfo  {journal}
  {Phys. Lett. B}\ }\textbf {\bibinfo {volume} {729}},\ \bibinfo {pages} {9}
  (\bibinfo {year} {2014})},\ \Eprint {http://arxiv.org/abs/1311.7552}
  {arXiv:1311.7552 [hep-ph]} \BibitemShut {NoStop}%
\bibitem [{\citenamefont {Pelaez}\ and\ \citenamefont
  {Rios}(2006)}]{Pelaez:2006nj}%
  \BibitemOpen
  \bibfield  {author} {\bibinfo {author} {\bibfnamefont {J.}~\bibnamefont
  {Pelaez}}\ and\ \bibinfo {author} {\bibfnamefont {G.}~\bibnamefont {Rios}},\
  }\href {\doibase 10.1103/PhysRevLett.97.242002} {\bibfield  {journal}
  {\bibinfo  {journal} {Phys. Rev. Lett.}\ }\textbf {\bibinfo {volume} {97}},\
  \bibinfo {pages} {242002} (\bibinfo {year} {2006})},\ \Eprint
  {http://arxiv.org/abs/hep-ph/0610397} {arXiv:hep-ph/0610397} \BibitemShut
  {NoStop}%
\bibitem [{\citenamefont {Nieves}\ \emph {et~al.}(2011)\citenamefont {Nieves},
  \citenamefont {Pich},\ and\ \citenamefont {Ruiz~Arriola}}]{Nieves:2011gb}%
  \BibitemOpen
  \bibfield  {author} {\bibinfo {author} {\bibfnamefont {J.}~\bibnamefont
  {Nieves}}, \bibinfo {author} {\bibfnamefont {A.}~\bibnamefont {Pich}}, \ and\
  \bibinfo {author} {\bibfnamefont {E.}~\bibnamefont {Ruiz~Arriola}},\ }\href
  {\doibase 10.1103/PhysRevD.84.096002} {\bibfield  {journal} {\bibinfo
  {journal} {Phys. Rev. D}\ }\textbf {\bibinfo {volume} {84}},\ \bibinfo
  {pages} {096002} (\bibinfo {year} {2011})},\ \Eprint
  {http://arxiv.org/abs/1107.3247} {arXiv:1107.3247 [hep-ph]} \BibitemShut
  {NoStop}%
\bibitem [{\citenamefont {Ruiz~de Elvira}\ \emph {et~al.}(2011)\citenamefont
  {Ruiz~de Elvira}, \citenamefont {Pelaez}, \citenamefont {Pennington},\ and\
  \citenamefont {Wilson}}]{RuizdeElvira:2010cs}%
  \BibitemOpen
  \bibfield  {author} {\bibinfo {author} {\bibfnamefont {J.}~\bibnamefont
  {Ruiz~de Elvira}}, \bibinfo {author} {\bibfnamefont {J.}~\bibnamefont
  {Pelaez}}, \bibinfo {author} {\bibfnamefont {M.}~\bibnamefont {Pennington}},
  \ and\ \bibinfo {author} {\bibfnamefont {D.}~\bibnamefont {Wilson}},\ }\href
  {\doibase 10.1103/PhysRevD.84.096006} {\bibfield  {journal} {\bibinfo
  {journal} {Phys. Rev. D}\ }\textbf {\bibinfo {volume} {84}},\ \bibinfo
  {pages} {096006} (\bibinfo {year} {2011})},\ \Eprint
  {http://arxiv.org/abs/1009.6204} {arXiv:1009.6204 [hep-ph]} \BibitemShut
  {NoStop}%
\bibitem [{\citenamefont {Prelovsek}\ and\ \citenamefont
  {Leskovec}(2013)}]{Prelovsek:2013cra}%
  \BibitemOpen
  \bibfield  {author} {\bibinfo {author} {\bibfnamefont {S.}~\bibnamefont
  {Prelovsek}}\ and\ \bibinfo {author} {\bibfnamefont {L.}~\bibnamefont
  {Leskovec}},\ }\href {\doibase 10.1103/PhysRevLett.111.192001} {\bibfield
  {journal} {\bibinfo  {journal} {Phys. Rev. Lett.}\ }\textbf {\bibinfo
  {volume} {111}},\ \bibinfo {pages} {192001} (\bibinfo {year} {2013})},\
  \Eprint {http://arxiv.org/abs/1307.5172} {arXiv:1307.5172 [hep-lat]}
  \BibitemShut {NoStop}%
%%CITATION = ARXIV:1307.5172;%%
\bibitem [{\citenamefont {Cotanch}\ and\ \citenamefont
  {Maris}(2002)}]{Cotanch:2002vj}%
  \BibitemOpen
  \bibfield  {author} {\bibinfo {author} {\bibfnamefont {S.~R.}\ \bibnamefont
  {Cotanch}}\ and\ \bibinfo {author} {\bibfnamefont {P.}~\bibnamefont
  {Maris}},\ }\href {\doibase 10.1103/PhysRevD.66.116010} {\bibfield  {journal}
  {\bibinfo  {journal} {Phys. Rev. D}\ }\textbf {\bibinfo {volume} {66}},\
  \bibinfo {pages} {116010} (\bibinfo {year} {2002})},\ \Eprint
  {http://arxiv.org/abs/hep-ph/0210151} {arXiv:hep-ph/0210151} \BibitemShut
  {NoStop}%
\bibitem [{\citenamefont {Chang}\ and\ \citenamefont
  {Roberts}(2009)}]{Chang:2009zb}%
  \BibitemOpen
  \bibfield  {author} {\bibinfo {author} {\bibfnamefont {L.}~\bibnamefont
  {Chang}}\ and\ \bibinfo {author} {\bibfnamefont {C.~D.}\ \bibnamefont
  {Roberts}},\ }\href {\doibase 10.1103/PhysRevLett.103.081601} {\bibfield
  {journal} {\bibinfo  {journal} {Phys. Rev. Lett.}\ }\textbf {\bibinfo
  {volume} {103}},\ \bibinfo {pages} {081601} (\bibinfo {year} {2009})},\
  \Eprint {http://arxiv.org/abs/0903.5461} {arXiv:0903.5461 [nucl-th]}
  \BibitemShut {NoStop}%
\bibitem [{\citenamefont {Fischer}\ and\ \citenamefont
  {Williams}(2009)}]{Fischer:2009jm}%
  \BibitemOpen
  \bibfield  {author} {\bibinfo {author} {\bibfnamefont {C.~S.}\ \bibnamefont
  {Fischer}}\ and\ \bibinfo {author} {\bibfnamefont {R.}~\bibnamefont
  {Williams}},\ }\href {\doibase 10.1103/PhysRevLett.103.122001} {\bibfield
  {journal} {\bibinfo  {journal} {Phys. Rev. Lett.}\ }\textbf {\bibinfo
  {volume} {103}},\ \bibinfo {pages} {122001} (\bibinfo {year} {2009})},\
  \Eprint {http://arxiv.org/abs/0905.2291} {arXiv:0905.2291 [hep-ph]}
  \BibitemShut {NoStop}%
\bibitem [{\citenamefont {Williams}\ \emph {et~al.}(2016)\citenamefont
  {Williams}, \citenamefont {Fischer},\ and\ \citenamefont
  {Heupel}}]{Williams:2015cvx}%
  \BibitemOpen
  \bibfield  {author} {\bibinfo {author} {\bibfnamefont {R.}~\bibnamefont
  {Williams}}, \bibinfo {author} {\bibfnamefont {C.~S.}\ \bibnamefont
  {Fischer}}, \ and\ \bibinfo {author} {\bibfnamefont {W.}~\bibnamefont
  {Heupel}},\ }\href {\doibase 10.1103/PhysRevD.93.034026} {\bibfield
  {journal} {\bibinfo  {journal} {Phys. Rev. D}\ }\textbf {\bibinfo {volume}
  {93}},\ \bibinfo {pages} {034026} (\bibinfo {year} {2016})},\ \Eprint
  {http://arxiv.org/abs/1512.00455} {arXiv:1512.00455 [hep-ph]} \BibitemShut
  {NoStop}%
\bibitem [{\citenamefont {Williams}(2019)}]{Williams:2018adr}%
  \BibitemOpen
  \bibfield  {author} {\bibinfo {author} {\bibfnamefont {R.}~\bibnamefont
  {Williams}},\ }\href {\doibase 10.1016/j.physletb.2019.134943} {\bibfield
  {journal} {\bibinfo  {journal} {Phys. Lett. B}\ }\textbf {\bibinfo {volume}
  {798}},\ \bibinfo {pages} {134943} (\bibinfo {year} {2019})},\ \Eprint
  {http://arxiv.org/abs/1804.11161} {arXiv:1804.11161 [hep-ph]} \BibitemShut
  {NoStop}%
\bibitem [{\citenamefont {Khvedelidze}\ and\ \citenamefont
  {Kvinikhidze}(1992)}]{Khvedelidze:1991qb}%
  \BibitemOpen
  \bibfield  {author} {\bibinfo {author} {\bibfnamefont {A.~M.}\ \bibnamefont
  {Khvedelidze}}\ and\ \bibinfo {author} {\bibfnamefont {A.~N.}\ \bibnamefont
  {Kvinikhidze}},\ }\href {\doibase 10.1007/BF01018820} {\bibfield  {journal}
  {\bibinfo  {journal} {Theor. Math. Phys.}\ }\textbf {\bibinfo {volume}
  {90}},\ \bibinfo {pages} {62} (\bibinfo {year} {1992})}\BibitemShut {NoStop}%
%%CITATION = TMPHA,90,62;%%
\bibitem [{\citenamefont {Eichmann}\ \emph
  {et~al.}(2016{\natexlab{b}})\citenamefont {Eichmann}, \citenamefont
  {Fischer},\ and\ \citenamefont {Sanchis-Alepuz}}]{Eichmann:2016hgl}%
  \BibitemOpen
  \bibfield  {author} {\bibinfo {author} {\bibfnamefont {G.}~\bibnamefont
  {Eichmann}}, \bibinfo {author} {\bibfnamefont {C.~S.}\ \bibnamefont
  {Fischer}}, \ and\ \bibinfo {author} {\bibfnamefont {H.}~\bibnamefont
  {Sanchis-Alepuz}},\ }\href {\doibase 10.1103/PhysRevD.94.094033} {\bibfield
  {journal} {\bibinfo  {journal} {Phys. Rev. D}\ }\textbf {\bibinfo {volume}
  {94}},\ \bibinfo {pages} {094033} (\bibinfo {year} {2016}{\natexlab{b}})},\
  \Eprint {http://arxiv.org/abs/1607.05748} {arXiv:1607.05748 [hep-ph]}
  \BibitemShut {NoStop}%
\bibitem [{\citenamefont {Eichmann}\ \emph
  {et~al.}(2016{\natexlab{c}})\citenamefont {Eichmann}, \citenamefont
  {Sanchis-Alepuz}, \citenamefont {Williams}, \citenamefont {Alkofer},\ and\
  \citenamefont {Fischer}}]{Eichmann:2016yit}%
  \BibitemOpen
  \bibfield  {author} {\bibinfo {author} {\bibfnamefont {G.}~\bibnamefont
  {Eichmann}}, \bibinfo {author} {\bibfnamefont {H.}~\bibnamefont
  {Sanchis-Alepuz}}, \bibinfo {author} {\bibfnamefont {R.}~\bibnamefont
  {Williams}}, \bibinfo {author} {\bibfnamefont {R.}~\bibnamefont {Alkofer}}, \
  and\ \bibinfo {author} {\bibfnamefont {C.~S.}\ \bibnamefont {Fischer}},\
  }\href {\doibase 10.1016/j.ppnp.2016.07.001} {\bibfield  {journal} {\bibinfo
  {journal} {Prog. Part. Nucl. Phys.}\ }\textbf {\bibinfo {volume} {91}},\
  \bibinfo {pages} {1} (\bibinfo {year} {2016}{\natexlab{c}})},\ \Eprint
  {http://arxiv.org/abs/1606.09602} {arXiv:1606.09602 [hep-ph]} \BibitemShut
  {NoStop}%
%%CITATION = ARXIV:1606.09602;%%
\bibitem [{\citenamefont {Sanchis-Alepuz}\ and\ \citenamefont
  {Williams}(2018)}]{Sanchis-Alepuz:2017jjd}%
  \BibitemOpen
  \bibfield  {author} {\bibinfo {author} {\bibfnamefont {H.}~\bibnamefont
  {Sanchis-Alepuz}}\ and\ \bibinfo {author} {\bibfnamefont {R.}~\bibnamefont
  {Williams}},\ }\href {\doibase 10.1016/j.cpc.2018.05.020} {\bibfield
  {journal} {\bibinfo  {journal} {Comput. Phys. Commun.}\ }\textbf {\bibinfo
  {volume} {232}},\ \bibinfo {pages} {1} (\bibinfo {year} {2018})},\ \Eprint
  {http://arxiv.org/abs/1710.04903} {arXiv:1710.04903 [hep-ph]} \BibitemShut
  {NoStop}%
%%CITATION = ARXIV:1710.04903;%%
\bibitem [{\citenamefont {Yokojima}\ \emph {et~al.}(1993)\citenamefont
  {Yokojima}, \citenamefont {Komachiya},\ and\ \citenamefont
  {Fukuda}}]{Yokojima:1993np}%
  \BibitemOpen
  \bibfield  {author} {\bibinfo {author} {\bibfnamefont {S.}~\bibnamefont
  {Yokojima}}, \bibinfo {author} {\bibfnamefont {M.}~\bibnamefont {Komachiya}},
  \ and\ \bibinfo {author} {\bibfnamefont {R.}~\bibnamefont {Fukuda}},\ }\href
  {\doibase 10.1016/0550-3213(93)90459-3} {\bibfield  {journal} {\bibinfo
  {journal} {Nucl. Phys. B}\ }\textbf {\bibinfo {volume} {390}},\ \bibinfo
  {pages} {319} (\bibinfo {year} {1993})}\BibitemShut {NoStop}%
\bibitem [{\citenamefont {Capstick}\ and\ \citenamefont
  {Roberts}(2000)}]{Capstick:2000qj}%
  \BibitemOpen
  \bibfield  {author} {\bibinfo {author} {\bibfnamefont {S.}~\bibnamefont
  {Capstick}}\ and\ \bibinfo {author} {\bibfnamefont {W.}~\bibnamefont
  {Roberts}},\ }\href {\doibase 10.1016/S0146-6410(00)00109-5} {\bibfield
  {journal} {\bibinfo  {journal} {Prog. Part. Nucl. Phys.}\ }\textbf {\bibinfo
  {volume} {45}},\ \bibinfo {pages} {S241} (\bibinfo {year} {2000})},\ \Eprint
  {http://arxiv.org/abs/nucl-th/0008028} {arXiv:nucl-th/0008028} \BibitemShut
  {NoStop}%
\bibitem [{\citenamefont {Kvinikhidze}\ and\ \citenamefont
  {Blankleider}(2014)}]{Kvinikhidze:2014yqa}%
  \BibitemOpen
  \bibfield  {author} {\bibinfo {author} {\bibfnamefont {A.}~\bibnamefont
  {Kvinikhidze}}\ and\ \bibinfo {author} {\bibfnamefont {B.}~\bibnamefont
  {Blankleider}},\ }\href {\doibase 10.1103/PhysRevD.90.045042} {\bibfield
  {journal} {\bibinfo  {journal} {Phys. Rev. D}\ }\textbf {\bibinfo {volume}
  {90}},\ \bibinfo {pages} {045042} (\bibinfo {year} {2014})},\ \Eprint
  {http://arxiv.org/abs/1406.5599} {arXiv:1406.5599 [hep-ph]} \BibitemShut
  {NoStop}%
\bibitem [{\citenamefont {Maris}\ and\ \citenamefont
  {Roberts}(2003)}]{Maris:2003vk}%
  \BibitemOpen
  \bibfield  {author} {\bibinfo {author} {\bibfnamefont {P.}~\bibnamefont
  {Maris}}\ and\ \bibinfo {author} {\bibfnamefont {C.~D.}\ \bibnamefont
  {Roberts}},\ }\href {\doibase 10.1142/S0218301303001326} {\bibfield
  {journal} {\bibinfo  {journal} {Int. J. Mod. Phys. E}\ }\textbf {\bibinfo
  {volume} {12}},\ \bibinfo {pages} {297} (\bibinfo {year} {2003})},\ \Eprint
  {http://arxiv.org/abs/nucl-th/0301049} {arXiv:nucl-th/0301049} \BibitemShut
  {NoStop}%
\bibitem [{\citenamefont {Wallbott}\ \emph {et~al.}(2019)\citenamefont
  {Wallbott}, \citenamefont {Eichmann},\ and\ \citenamefont
  {Fischer}}]{Wallbott:2019dng}%
  \BibitemOpen
  \bibfield  {author} {\bibinfo {author} {\bibfnamefont {P.~C.}\ \bibnamefont
  {Wallbott}}, \bibinfo {author} {\bibfnamefont {G.}~\bibnamefont {Eichmann}},
  \ and\ \bibinfo {author} {\bibfnamefont {C.~S.}\ \bibnamefont {Fischer}},\
  }\href {\doibase 10.1103/PhysRevD.100.014033} {\bibfield  {journal} {\bibinfo
   {journal} {Phys. Rev.}\ }\textbf {\bibinfo {volume} {D100}},\ \bibinfo
  {pages} {014033} (\bibinfo {year} {2019})},\ \Eprint
  {http://arxiv.org/abs/1905.02615} {arXiv:1905.02615 [hep-ph]} \BibitemShut
  {NoStop}%
%%CITATION = ARXIV:1905.02615;%%
\bibitem [{\citenamefont {Wallbott}\ \emph {et~al.}(2020)\citenamefont
  {Wallbott}, \citenamefont {Eichmann},\ and\ \citenamefont
  {Fischer}}]{Wallbott:2020jzh}%
  \BibitemOpen
  \bibfield  {author} {\bibinfo {author} {\bibfnamefont {P.~C.}\ \bibnamefont
  {Wallbott}}, \bibinfo {author} {\bibfnamefont {G.}~\bibnamefont {Eichmann}},
  \ and\ \bibinfo {author} {\bibfnamefont {C.~S.}\ \bibnamefont {Fischer}},\
  }\href@noop {} {\  (\bibinfo {year} {2020})},\ \Eprint
  {http://arxiv.org/abs/2003.12407} {arXiv:2003.12407 [hep-ph]} \BibitemShut
  {NoStop}%
\bibitem [{\citenamefont {Maris}\ and\ \citenamefont
  {Tandy}(1999)}]{Maris:1999nt}%
  \BibitemOpen
  \bibfield  {author} {\bibinfo {author} {\bibfnamefont {P.}~\bibnamefont
  {Maris}}\ and\ \bibinfo {author} {\bibfnamefont {P.~C.}\ \bibnamefont
  {Tandy}},\ }\href {\doibase 10.1103/PhysRevC.60.055214} {\bibfield  {journal}
  {\bibinfo  {journal} {Phys. Rev. C}\ }\textbf {\bibinfo {volume} {60}},\
  \bibinfo {pages} {055214} (\bibinfo {year} {1999})},\ \Eprint
  {http://arxiv.org/abs/nucl-th/9905056} {arXiv:nucl-th/9905056} \BibitemShut
  {NoStop}%
\bibitem [{\citenamefont {Watson}\ \emph {et~al.}(2004)\citenamefont {Watson},
  \citenamefont {Cassing},\ and\ \citenamefont {Tandy}}]{Watson:2004kd}%
  \BibitemOpen
  \bibfield  {author} {\bibinfo {author} {\bibfnamefont {P.}~\bibnamefont
  {Watson}}, \bibinfo {author} {\bibfnamefont {W.}~\bibnamefont {Cassing}}, \
  and\ \bibinfo {author} {\bibfnamefont {P.}~\bibnamefont {Tandy}},\ }\href
  {\doibase 10.1007/s00601-004-0067-x} {\bibfield  {journal} {\bibinfo
  {journal} {Few Body Syst.}\ }\textbf {\bibinfo {volume} {35}},\ \bibinfo
  {pages} {129} (\bibinfo {year} {2004})},\ \Eprint
  {http://arxiv.org/abs/hep-ph/0406340} {arXiv:hep-ph/0406340} \BibitemShut
  {NoStop}%
\bibitem [{\citenamefont {Watson}\ and\ \citenamefont
  {Cassing}(2004)}]{Watson:2004jq}%
  \BibitemOpen
  \bibfield  {author} {\bibinfo {author} {\bibfnamefont {P.}~\bibnamefont
  {Watson}}\ and\ \bibinfo {author} {\bibfnamefont {W.}~\bibnamefont
  {Cassing}},\ }\href {\doibase 10.1007/s00601-004-0063-1} {\bibfield
  {journal} {\bibinfo  {journal} {Few Body Syst.}\ }\textbf {\bibinfo {volume}
  {35}},\ \bibinfo {pages} {99} (\bibinfo {year} {2004})},\ \Eprint
  {http://arxiv.org/abs/hep-ph/0405287} {arXiv:hep-ph/0405287} \BibitemShut
  {NoStop}%
\bibitem [{\citenamefont {Miramontes}\ and\ \citenamefont
  {Sanchis-Alepuz}(2019)}]{Miramontes:2019mco}%
  \BibitemOpen
  \bibfield  {author} {\bibinfo {author} {\bibfnamefont {A.~S.}\ \bibnamefont
  {Miramontes}}\ and\ \bibinfo {author} {\bibfnamefont {H.}~\bibnamefont
  {Sanchis-Alepuz}},\ }\href {\doibase 10.1140/epja/i2019-12847-6} {\bibfield
  {journal} {\bibinfo  {journal} {Eur. Phys. J. A}\ }\textbf {\bibinfo {volume}
  {55}},\ \bibinfo {pages} {170} (\bibinfo {year} {2019})},\ \Eprint
  {http://arxiv.org/abs/1906.06227} {arXiv:1906.06227 [hep-ph]} \BibitemShut
  {NoStop}%
\bibitem [{\citenamefont {Tripolt}\ \emph {et~al.}(2017)\citenamefont
  {Tripolt}, \citenamefont {Haritan}, \citenamefont {Wambach},\ and\
  \citenamefont {Moiseyev}}]{Tripolt:2016cya}%
  \BibitemOpen
  \bibfield  {author} {\bibinfo {author} {\bibfnamefont {R.-A.}\ \bibnamefont
  {Tripolt}}, \bibinfo {author} {\bibfnamefont {I.}~\bibnamefont {Haritan}},
  \bibinfo {author} {\bibfnamefont {J.}~\bibnamefont {Wambach}}, \ and\
  \bibinfo {author} {\bibfnamefont {N.}~\bibnamefont {Moiseyev}},\ }\href
  {\doibase 10.1016/j.physletb.2017.10.001} {\bibfield  {journal} {\bibinfo
  {journal} {Phys. Lett. B}\ }\textbf {\bibinfo {volume} {774}},\ \bibinfo
  {pages} {411} (\bibinfo {year} {2017})},\ \Eprint
  {http://arxiv.org/abs/1610.03252} {arXiv:1610.03252 [hep-ph]} \BibitemShut
  {NoStop}%
\bibitem [{\citenamefont {Tripolt}\ \emph {et~al.}(2019)\citenamefont
  {Tripolt}, \citenamefont {Gubler}, \citenamefont {Ulybyshev},\ and\
  \citenamefont {Von~Smekal}}]{Tripolt:2018xeo}%
  \BibitemOpen
  \bibfield  {author} {\bibinfo {author} {\bibfnamefont {R.-A.}\ \bibnamefont
  {Tripolt}}, \bibinfo {author} {\bibfnamefont {P.}~\bibnamefont {Gubler}},
  \bibinfo {author} {\bibfnamefont {M.}~\bibnamefont {Ulybyshev}}, \ and\
  \bibinfo {author} {\bibfnamefont {L.}~\bibnamefont {Von~Smekal}},\ }\href
  {\doibase 10.1016/j.cpc.2018.11.012} {\bibfield  {journal} {\bibinfo
  {journal} {Comput. Phys. Commun.}\ }\textbf {\bibinfo {volume} {237}},\
  \bibinfo {pages} {129} (\bibinfo {year} {2019})},\ \Eprint
  {http://arxiv.org/abs/1801.10348} {arXiv:1801.10348 [hep-ph]} \BibitemShut
  {NoStop}%
\bibitem [{\citenamefont {Binosi}\ and\ \citenamefont
  {Tripolt}(2020)}]{Binosi:2019ecz}%
  \BibitemOpen
  \bibfield  {author} {\bibinfo {author} {\bibfnamefont {D.}~\bibnamefont
  {Binosi}}\ and\ \bibinfo {author} {\bibfnamefont {R.-A.}\ \bibnamefont
  {Tripolt}},\ }\href {\doibase 10.1016/j.physletb.2019.135171} {\bibfield
  {journal} {\bibinfo  {journal} {Phys. Lett. B}\ }\textbf {\bibinfo {volume}
  {801}},\ \bibinfo {pages} {135171} (\bibinfo {year} {2020})},\ \Eprint
  {http://arxiv.org/abs/1904.08172} {arXiv:1904.08172 [hep-ph]} \BibitemShut
  {NoStop}%
\bibitem [{\citenamefont {Eichmann}\ \emph {et~al.}(2019)\citenamefont
  {Eichmann}, \citenamefont {Duarte}, \citenamefont {Peña},\ and\
  \citenamefont {Stadler}}]{Eichmann:2019dts}%
  \BibitemOpen
  \bibfield  {author} {\bibinfo {author} {\bibfnamefont {G.}~\bibnamefont
  {Eichmann}}, \bibinfo {author} {\bibfnamefont {P.}~\bibnamefont {Duarte}},
  \bibinfo {author} {\bibfnamefont {M.}~\bibnamefont {Peña}}, \ and\ \bibinfo
  {author} {\bibfnamefont {A.}~\bibnamefont {Stadler}},\ }\href {\doibase
  10.1103/PhysRevD.100.094001} {\bibfield  {journal} {\bibinfo  {journal}
  {Phys. Rev. D}\ }\textbf {\bibinfo {volume} {100}},\ \bibinfo {pages}
  {094001} (\bibinfo {year} {2019})},\ \Eprint
  {http://arxiv.org/abs/1907.05402} {arXiv:1907.05402 [hep-ph]} \BibitemShut
  {NoStop}%
\bibitem [{\citenamefont {Mader}\ \emph {et~al.}(2011)\citenamefont {Mader},
  \citenamefont {Eichmann}, \citenamefont {Blank},\ and\ \citenamefont
  {Krassnigg}}]{Mader:2011zf}%
  \BibitemOpen
  \bibfield  {author} {\bibinfo {author} {\bibfnamefont {V.}~\bibnamefont
  {Mader}}, \bibinfo {author} {\bibfnamefont {G.}~\bibnamefont {Eichmann}},
  \bibinfo {author} {\bibfnamefont {M.}~\bibnamefont {Blank}}, \ and\ \bibinfo
  {author} {\bibfnamefont {A.}~\bibnamefont {Krassnigg}},\ }\href {\doibase
  10.1103/PhysRevD.84.034012} {\bibfield  {journal} {\bibinfo  {journal} {Phys.
  Rev. D}\ }\textbf {\bibinfo {volume} {84}},\ \bibinfo {pages} {034012}
  (\bibinfo {year} {2011})},\ \Eprint {http://arxiv.org/abs/1106.3159}
  {arXiv:1106.3159 [hep-ph]} \BibitemShut {NoStop}%
\end{thebibliography}%

\end{document}